\documentclass[preprint,12pt, sort&compress,numbers]{elsarticle}

\usepackage{soul}
\usepackage{amssymb, amsmath}
\usepackage{stmaryrd}

\usepackage{hyperref}

\usepackage{comment}
\usepackage{arydshln}
\usepackage{bm}
\usepackage[textsize=tiny,disable]{todonotes}

\usepackage{tikz}
\usetikzlibrary{positioning}
\usepackage{pgfplots}
\pgfplotsset{compat=1.10}
\usepgfplotslibrary{groupplots,fillbetween}
\usetikzlibrary{spy, backgrounds}
\usetikzlibrary{shapes.geometric, arrows.meta}

\newcommand{\cO}{\mathcal{O}}

\newcommand{\cL}{\mathcal{L}}

\newcommand{\cN}{\mathcal{N}}
\newcommand{\cT}{\mathcal{T}}

\newcommand{\rR}{\mathbb{R}}
\newcommand{\rI}{\mathbb{I}}
\newcommand{\rQ}{\mathbb{Q}}
\newcommand{\rU}{\mathbb{U}}
\newcommand{\rP}{\mathbb{P}}
\newcommand{\rC}{\mathbb{C}}

\newcommand{\bA}{\mathbf{A}}
\newcommand{\bB}{\mathbf{B}}

\newcommand{\bL}{\mathbf{L}}
\newcommand{\bS}{\mathbf{S}}
\newcommand{\bV}{\mathbf{V}}
\newcommand{\bP}{\mathbf{P}}
\newcommand{\bM}{\mathbf{M}}
\newcommand{\bQ}{\mathbf{Q}}
\newcommand{\bF}{\mathbf{F}}
\newcommand{\bG}{\mathbf{G}}
\newcommand{\bK}{\mathbf{K}}
\newcommand{\bI}{\mathbf{I}}
\newcommand{\bk}{\mathbf{k}}
\newcommand{\bn}{\mathbf{n}}
\newcommand{\bq}{\mathbf{q}}
\newcommand{\bx}{\mathbf{x}}
\newcommand{\bu}{\mathbf{u}}
\newcommand{\bv}{\mathbf{v}}
\newcommand{\bus}{\bu^{\dagger}}
\newcommand{\bg}{\mathbf{g}}
\newcommand{\tbq}{\tilde{\bq}}
\newcommand{\bqs}{\mathbf{q}^{\dagger}}

\newcommand{\f}{\mathbf{f}}
\newcommand{\bPhi}{\mathbf{\Phi}}
\newcommand{\bphi}{\bm{\phi}}

\newcommand{\ps}{p^{\dagger}}

\newcommand{\hL}{{\hat{L}}}
\newcommand{\hB}{{\hat{B}}}

\newcommand{\hbq}{{\hat{\bq}}}
\newcommand{\hbqs}{{\hat{\bq}^{\dagger}}}
\newcommand{\hbB}{{\hat{\bB}}}

\newcommand{\hbL}{{\hat{\bL}}}

\newcommand{\us}{u^{\dagger}}

\newcommand{\bOmega}{\overline{\Omega}}

\def\inprod#1#2{\left\langle #1, #2 \right\rangle}
\def\form#1#2#3{#1\!\inprod{#2}{#3}}
\def\form#1#2#3{#1\!\inprod{#2}{#3}}
\def\form#1#2#3{#1\!\inprod{#2}{#3}}

\def\jump#1{\left\llbracket #1 \right\rrbracket}
\def\avg#1{\left\{\!\!\left\{ #1 \right\}\!\!\right\}}

\newcommand{\logLogSlopeTriangle}[6]
{

    \pgfplotsextra
    {
        \pgfkeysgetvalue{/pgfplots/xmin}{\xmin}
        \pgfkeysgetvalue{/pgfplots/xmax}{\xmax}
        \pgfkeysgetvalue{/pgfplots/ymin}{\ymin}
        \pgfkeysgetvalue{/pgfplots/ymax}{\ymax}

        \pgfmathsetmacro{\xArel}{#1}
        \pgfmathsetmacro{\yArel}{#3}
        \pgfmathsetmacro{\xBrel}{#1-#2}
        \pgfmathsetmacro{\yBrel}{\yArel}
        \pgfmathsetmacro{\xCrel}{\xArel}

        \pgfmathsetmacro{\lnxB}{\xmin*(1-(#1-#2))+\xmax*(#1-#2)} 
        \pgfmathsetmacro{\lnxA}{\xmin*(1-#1)+\xmax*#1} 
        \pgfmathsetmacro{\lnyA}{\ymin*(1-#3)+\ymax*#3} 
        \pgfmathsetmacro{\lnyC}{\lnyA+#4*(\lnxA-\lnxB)}
        \pgfmathsetmacro{\yCrel}{(\lnyC-\ymin)/(\ymax-\ymin)} 

        \coordinate (A) at (rel axis cs:\xArel,\yArel);
        \coordinate (B) at (rel axis cs:\xBrel,\yBrel);
        \coordinate (C) at (rel axis cs:\xCrel,\yCrel);

        \draw[#5]   (A)-- node[pos=0.5,anchor=#6] {1}
                    (B)-- 
                    (C)-- node[pos=0.5,anchor=west] {#4}
                    cycle;
    }
}

\newcommand{\logLogReverseSlopeTriangle}[6]
{

    \pgfplotsextra
    {
        \pgfkeysgetvalue{/pgfplots/xmin}{\xmin}
        \pgfkeysgetvalue{/pgfplots/xmax}{\xmax}
        \pgfkeysgetvalue{/pgfplots/ymin}{\ymin}
        \pgfkeysgetvalue{/pgfplots/ymax}{\ymax}

        \pgfmathsetmacro{\xArel}{#1}
        \pgfmathsetmacro{\yArel}{#3}
        \pgfmathsetmacro{\xBrel}{#1+#2}
        \pgfmathsetmacro{\yBrel}{\yArel}
        \pgfmathsetmacro{\xCrel}{\xArel}

        \pgfmathsetmacro{\lnxB}{\xmin*(1-(#1-#2))+\xmax*(#1-#2)} 
        \pgfmathsetmacro{\lnxA}{\xmin*(1-#1)+\xmax*#1} 
        \pgfmathsetmacro{\lnyA}{\ymin*(1-#3)+\ymax*#3} 
        \pgfmathsetmacro{\lnyC}{\lnyA-#4*(\lnxA-\lnxB)}
        \pgfmathsetmacro{\yCrel}{(\lnyC-\ymin)/(\ymax-\ymin)} 

        \coordinate (A) at (rel axis cs:\xArel,\yArel);
        \coordinate (B) at (rel axis cs:\xBrel,\yBrel);
        \coordinate (C) at (rel axis cs:\xCrel,\yCrel);

        \draw[#5]   (A)-- node[pos=0.5,anchor=#6] {1}
                    (B)-- 
                    (C)-- node[pos=0.5,anchor=east] {#4}
                    cycle;
    }
}

\pgfplotsset{select coords between index/.style 2 args={
    x filter/.code={
        \ifnum\coordindex<#1\fi
        \ifnum\coordindex>#2\fi
    }
}}

\newif\ifcommentcoloring
\commentcoloringfalse

\newcommand{\YC}[1]{\ifcommentcoloring \textcolor{red}{#1}\else #1\fi}
\newcommand{\PR}[1]{\ifcommentcoloring \textcolor{blue}{#1}\else #1\fi}
\newcommand{\KC}[1]{\ifcommentcoloring \textcolor{brown!70!black}{#1}\else #1\fi}
\newcommand{\TYL}[1]{\ifcommentcoloring \textcolor{magenta}{#1}\else #1\fi}

\journal{arXiv}

\begin{document}

\begin{frontmatter}



\title{
Train Small, Model Big: Scalable Physics Simulators via Reduced Order Modeling and Domain Decomposition
}


\author[med]{Seung Whan Chung\corref{cor1}}
\cortext[cor1]{corresponding author}
\ead{chung28@llnl.gov}
\author[casc]{Youngsoo Choi}
\author[aeed]{Pratanu Roy}
\author[qut]{Thomas Moore}
\author[ced]{Thomas Roy}
\author[ced]{Tiras Y. Lin}
\author[med]{Du T. Nguyen}
\author[msd,leaf]{Christopher Hahn}
\author[med,cemm]{Eric B. Duoss}
\author[msd]{Sarah E. Baker}

\affiliation[med]{organization={Material Engineering Division, Lawrence Livermore National Laboratory},
  city={Livermore},
  postcode={94550}, 
  state={CA},
  country={US}}
\affiliation[casc]{organization={Center for Applied Scientific Computing, Lawrence Livermore National Laboratory},
  city={Livermore},
  postcode={94550}, 
  state={CA},
  country={US}}
\affiliation[aeed]{organization={Atmospheric, Earth and Energy Division, Lawrence Livermore National Laboratory},
  city={Livermore},
  postcode={94550}, 
  state={CA},
  country={US}}
\affiliation[qut]{organization={School of Mechanical, Medical and Process Engineering, Queensland University of Technology}, 
  city={Brisbane}, 
  postcode={4000},
  state={QLD},
  country={Australia}
}
\affiliation[ced]{organization={Computational Engineering Division, Lawrence Livermore National Laboratory},
  city={Livermore},
  postcode={94550}, 
  state={CA},
  country={US}}
\affiliation[msd]{organization={Material Science Division, Lawrence Livermore National Laboratory},
  city={Livermore},
  postcode={94550}, 
  state={CA},
  country={US}}
\affiliation[leaf]{organization={Laboratory for Energy Applications for the Future, Lawrence Livermore National Laboratory},
  city={Livermore},
  postcode={94550}, 
  state={CA},
  country={US}}
\affiliation[cemm]{organization={Center for Engineered Materials, Manufacturing and Optimization, Lawrence Livermore National Laboratory},
  city={Livermore},
  postcode={94550}, 
  state={CA},
  country={US}}

\begin{abstract}
Numerous cutting-edge scientific technologies originate at the laboratory scale,
but transitioning them to practical industry applications is a formidable challenge.
Traditional pilot projects at intermediate scales are costly and time-consuming.
An alternative, the E-pilot, relies on high-fidelity numerical simulations,
but even these simulations can be computationally prohibitive at larger scales.
To \PR{overcome these limitations}, we propose a scalable, physics-constrained reduced order model (ROM) method.
ROM identifies critical physics modes from small-scale unit components,
projecting governing equations onto these modes to create a reduced model that retains essential physics details.
We also employ Discontinuous Galerkin Domain Decomposition (DG-DD) to apply ROM to unit components and interfaces, 
enabling the construction of large-scale global systems without data at such large scales.
This method is demonstrated on the Poisson and Stokes flow equations,
showing that it can solve equations about \PR{$15 - 40$} times faster with only  \PR{$\sim$} $1\%$ relative error.
Furthermore, 
\KC{ROM takes one order of magnitude less memory than the full order model,
enabling larger scale predictions at a given memory limitation.}
\end{abstract}




\end{frontmatter}


\section{Introduction}

The scale-up of nascent technologies is a notoriously difficult task.
While thousands of novel technologies are proposed in the academic literature each year,
only a small fraction leave the laboratory and find commercial success.
Among those that \textit{do} succeed,
the average time from conception to commercialization is \PR{as high as} 35 years~\cite{Gross2018}.
New industrial technologies typically require demonstration via a pilot plant,
which often cost tens of millions of dollars and can take years to design, construct and operate.
Given the urgent threat of climate change and the need for rapid decarbonization across the industrial sector,
approaches for accelerating technology scale-up are desperately needed.
\todo{KC: The last sentence would make sense only if we directly tackle the carbon-capture problem. Should we omit this sentence?}
\par
One promising approach is the development of ``E-pilots'':\todo{KC: removed digital twin. The concept we introduce here does not have its physical twin.}
fully resolved computational simulations that model industrial processes rigorously from first principles.
High fidelity simulations can often be constructed via conventional numerical methods
providing accurate predictions based on the underlying physics~\cite{Toselli2002,Cockburn2002,Wagner2001,pozrikidis1992boundary}.
However, fully resolved three-dimensional simulations of multiphysics problems at industrial scales often prove numerically infeasible.
As a concrete example, Singh~\textit{et al.}~\cite{Singh2022} developed a rigorous simulation of two-phase fluid flow
within a novel packing material for carbon capture.
While their simulations showed excellent agreement with experiment, they were expensive even at the laboratory scale:
simulating 30 seconds of unsteady two-phase flow within a 15~cm tall reactor required 3 days on 144 cores.
By way of comparison, Petra Nova's industrial-scale CCS absorber~\cite{petranova} is roughly $\sim750\times$ taller
and has about 10 million times more volume than the lab-scale system.
Clearly, high-fidelity simulations of such systems are numerically infeasible when conducted via standard methods.
\par
Our goal in this work is to develop a reduced modeling framework that can provide sufficiently accurate,
robust predictions at large scales in an efficient manner. It is worth taking a moment to expand on these goals.
By \textit{sufficiently accurate}, we mean simulations whose predictions match a rigorous numerical simulation to within a few percent relative error.
Engineering parameters such as material properties or system geometry come with similar degrees of uncertainty,
so further increasing mathematical accuracy beyond this point gives diminishing returns at the cost of increased computational time.
By \textit{robust}, we mean an ability to simulate over a wide range of parameters without catastrophic failure,
and in particular the ability to accurately predict \textit{emergent} properties
that arise as the size of the system grows from the laboratory scale to industrial scales.
For example, bench-scale reactors tend to be more isothermal (due to their relatively high specific surface area)
while industrial scale reactors are more susceptible to large temperature changes, which can then impact reaction conversion and selectivities.
\todo{Kevin: can we add a citation for this example?}
A challenge in developing this framework is that if a system is too large to simulate via conventional methods, then, by definition,
there is no numerical, system-scale data available with which to develop a data-based simulation approach.
A successful, data-driven framework must instead be trained on data on smaller domains,
and must then somehow extrapolate this training in a robust manner to larger scales where data \PR{are} unavailable.
\par
Data-driven machine learning techniques have been emerging
as a promising alternative to the conventional physics simulations~\cite{Brunton2016,Raissi2019,Li2020}.
Relying on a large amount of experimental/simulation data,
they attempt to replace the conventional physics simulation
with a sparse dynamical system~\cite{Brunton2016} or a neural network~\cite{Raissi2019,Li2020}.
However, these approaches have yet to be improved in their accuracy, \todo{KC: Youngsoo, do you happen to know any reference we can add here to support this sentence?}
particularly when the prediction is made outside the training data.
More fundamentally, even with a well-trained model at a particular scale,
it is not so clear how to extend it toward larger scales without any available data.
Such extrapolation in scale is the major bottleneck
in applying these data-driven methods directly as E-pilots.
\par
Projection-based reduced order model (ROM) can be a promising alternative for this type of problem,
taking the advantages of both the high-fidelity simulation and the data-driven approach.
It effectively approximates the solution with the most dominant physics `modes'
as given by proper orthogonal decomposition (POD) of the solution sample
data~\cite{choi2019space, choi2021space, choi2020gradient, copeland2022reduced, cheung2023local,
kim2022fast, carlberg2018conservative, kim2021efficient}.
Typically the first $10\sim20$ modes are sufficient to accurately represent the solution.
By projecting the PDEs onto a subspace spanned by these critical modes,
the number of unknowns can be reduced from thousands (as in a typical FEM discretization) to dozens.
Then, significant acceleration can be achieved by neglecting the minor details.
At the same time, ROMs can make robust extrapolative predictions
as they are built upon the high-fidelity simulation models.
\par
ROM has been widely used in many scientific and engineering applications, producing many variants.
Among them, component ROMs exhibit remarkable capability in extending
outcomes for large-scale systems based solely on data from smaller unit cells.
These component ROMs have been successfully employed across a spectrum of
physical problems, including lattice-type structure design optimization
\cite{mcbane2021component, mcbane2022stress},
microtruss structures \cite{eftang2014port}, heat
conduction/exchange \cite{huynh2013static, smetana2015new, vallaghe2014static},
linear elasticity \cite{huynh2013static}, and acoustic Helmholtz problems
\cite{huynh2013staticII}.  However, it is important to note that the
performance of these component approaches hinges on their ability to
accurately represent interfaces with POD bases separate from the interior of the domain.
This can add complexity to the decomposition and recomposition of domains with these component ROM methods.
\par
To address this challenge, we propose a novel physics-constrained
data-driven component ROM that offers both speed
and accuracy in predicting outcomes for large-scale systems, all without the
need to handle interfaces separately. To accomplish this, we leverage a
discontinuous-Galerkin domain decomposition (DG-DD) approach~\cite{Hansbo2005,Toselli2002}. 
The component-level ROM begins by identifying the dominant modes within a given
physics system based on component-level sample data. Subsequently, the
governing equation at component level is projected onto
the linear subspace defined by these identified modes, yielding a
component-level ROM.  DG-DD, on the other hand, decomposes a
larger global system into smaller-scale components, where the component-level
ROMs serve as foundational building blocks that are seamlessly integrated into
the overarching global-scale ROM.
In effect, we train our model on a small domain, and then compose many small domains together into a larger system.
Because the underlying ROM still solves the governing physics equations (albeit on a small function space with an optimally-chosen basis)
the methodology is robust to extrapolation, and in particular may be expected to capture `emergent phenomena',
such as large changes in temperature, not seen in the laboratory scale.
This integrated approach allows for accurate
and efficient predictions on a large scale, all while circumventing the need
of handling interfaces and the interiors of domains separately.
\par
There have been similar model reduction approaches that employ the discontinuous Galerkin projection.
For example,
the discontinuous Galerkin reduced basis element method
generates homogenous or particular solutions with parametrized forcing or boundary condition at subdomains,
and formulates the discontinuous Galerkin projection with them as spectral basis~\cite{Antonietti2016,Pacciarini2016}.
Their basis generation approach is distinctly different from our data-driven ROM,
in that \YC{our} ROM basis can be constructed from general samples, whether they are homogenous or particular, or a combination of both.
Similarly, Yano~\cite{Yano2019dg} applied ROM with empirical quadrature points for DG discretization of nonlinear conservation laws.
However, in this work the use of DG discretization is mainly to inherit the energy stability of the DG scheme,
and the domain decomposition is not taken into account.
\KC{Any of these works does not consider the potential of ROM and DG-DD for the robust and simple extrapolation in scale}\YC{, while our proposed method does.}
\par
In this paper, we focus on general linear PDE systems,
with the examples of the Poisson equation and the Stokes flow equation.
Our framework \KC{could allow} the possibility of extending to nonlinear systems,
\KC{by being} augmented with nonlinear ROM techniques such as the tensorial approach~\cite{Lassila2014} and hyper-reduction~\cite{Willcox2006,Chaturantabut2010,Yano2019,lauzon2022s}.
\par
The rest of the paper is organized as follows.
In Section~\ref{sec:formulation}, the proposed component reduced-order modeling is formulated
with respect to a general system of partial differential equations (PDE).
Then Section~\ref{sec:poisson}~and~\ref{sec:stokes} demonstrates the application of the general framework in Section~\ref{sec:formulation}
to the Poisson equation and the Stokes flow equation, respectively.

\section{General framework}\label{sec:formulation}
\todo[inline]{KC: removed the nonlinear terms now.}
\subsection{Physics governing equation and its domain decomposition}

We consider a generic form of a linear governing equation for a physics system state $\tbq$
on a two- or three-dimensional global domain $\Omega$,
\begin{equation}\label{eq:gov}
    \cL[\tbq] = \f,
\end{equation}
where $\cL$ \PR{is} the linear operator of $\tbq$.
Unless otherwise stated, we consider a time-independent solution $\tbq\in H^1(\Omega)^k$ with solution dimension $k$,
and $\f\in L_2(\Omega)^k$. However, the proposed model reduction method \PR{can readily be extended} to time-dependent problems as well.
\par
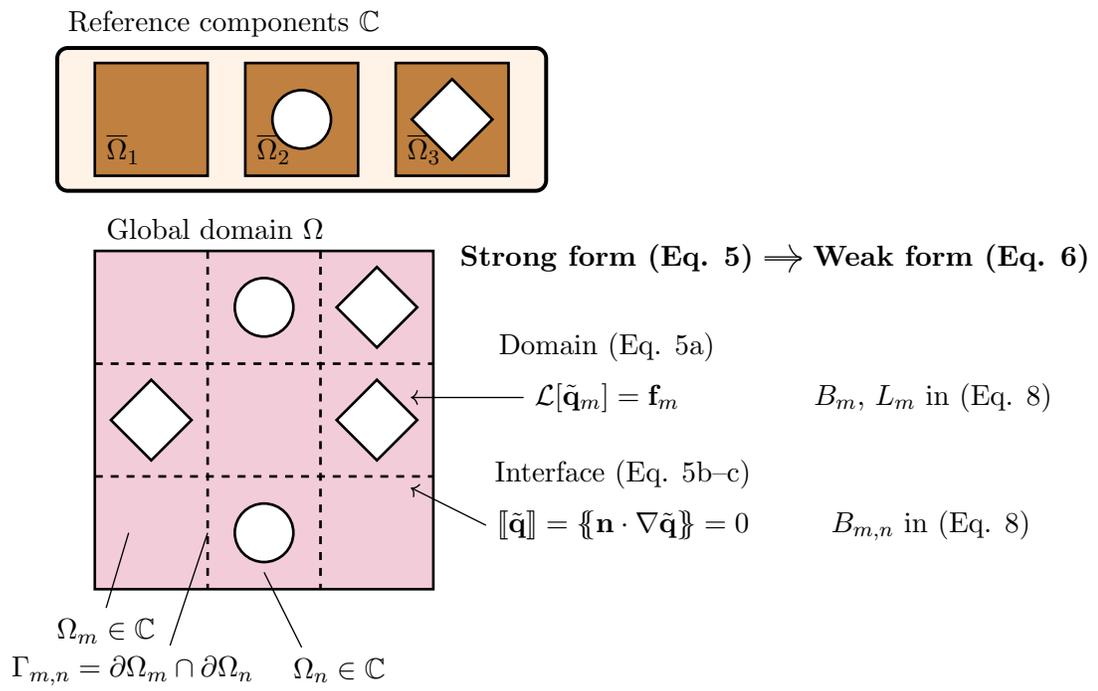
\begin{figure}[tbph]
    \begin{tikzpicture}[font=\small,]

\def\compsize{1.5}
\def\compgap{2.0}

\draw[rounded corners, draw=black, fill=orange!10, line width=1.5] (-0.5, -0.2) rectangle (6.0, 1.7);
\node[anchor=south west,] at (-0.5, 1.7) {Reference components $\rC$};

\draw[draw=black, fill=brown, line width=1.0,] (0, 0) node[anchor=south west,] {$\bOmega_1$} rectangle ++ (\compsize, \compsize);

\draw[draw=black, fill=brown, line width=1.0,] (\compgap, 0) node[anchor=south west,] {$\bOmega_2$} rectangle ++ (\compsize, \compsize);
\node[draw=black, fill=white, line width=1.0, circle, scale=2.,] at (\compgap+0.5*\compsize, 0.5*\compsize) {}; 

\draw[draw=black, fill=brown, line width=1.0,] (2*\compgap, 0) node[anchor=south west,] {$\bOmega_3$} rectangle ++ (\compsize, \compsize);
\node[draw=black, fill=white, line width=1.0, diamond, scale=2.,] at (2*\compgap+0.5*\compsize, 0.5*\compsize) {}; 

\draw[draw=black, fill=purple!20, line width=1.0,] (0, -1)
    node[anchor=south west,] {Global domain $\Omega$} rectangle ++(3*\compsize,-3*\compsize);
\draw[draw=black, dashed, line width=1.0,] (\compsize, -1) -- ++(0, -3*\compsize);
\draw[draw=black, dashed, line width=1.0,] (2*\compsize, -1) -- ++(0, -3*\compsize);
\draw[draw=black, dashed, line width=1.0,] (0, -1-\compsize) -- ++(3*\compsize, 0);
\draw[draw=black, dashed, line width=1.0,] (0, -1-2*\compsize) -- ++(3*\compsize, 0);

\node[draw=black, fill=white, line width=1.0, circle, scale=2.,] at (0 + 1.5*\compsize, -1 - 0.5*\compsize) {};
\node[draw=black, fill=white, line width=1.0, circle, scale=2.,] at (0 + 1.5*\compsize, -1 - 2.5*\compsize) {};
\node[draw=black, fill=white, line width=1.0, diamond, scale=2.,] at (0 + 2.5*\compsize, -1 - 0.5*\compsize) {};
\node[draw=black, fill=white, line width=1.0, diamond, scale=2.,] at (0 + 2.5*\compsize, -1 - 1.5*\compsize) {};
\node[draw=black, fill=white, line width=1.0, diamond, scale=2.,] at (0 + 0.5*\compsize, -1 - 1.5*\compsize) {};

\draw[draw=black, line width=0.5,] (0 + 1.5*\compsize, -1 - 2.85*\compsize) -- ++(0.5, -1.0)
    node[anchor=north, xshift=.5cm,] {$\Omega_n\in\rC$};
\draw[draw=black, line width=0.5,] (0 + 0.3*\compsize, -1 - 2.5*\compsize) -- ++(-0.3, -1.0)
    node[anchor=north, xshift=.0cm,] {$\Omega_m\in\rC$};
\draw[draw=black, line width=0.5,] (0 + 1*\compsize, -1 - 2.5*\compsize) -- ++(-0.5, -1.5)
    node[anchor=north, xshift=-.5cm,] {$\Gamma_{m,n}=\partial\Omega_m\cap\partial\Omega_n$};

\draw[<-, line width=0.5,] (0 + 2.8*\compsize, -1 - 1.3*\compsize) -- ++(1.5, 0)
    node[anchor=west,] (domstreq) {$\displaystyle\cL[\tbq_m] = \f_m$};
\node[anchor=south,] (domstreq1) at (domstreq.north) {Domain (Eq. \ref{eq:dd-gov}a)};
\node[anchor=south, yshift=0.5cm,] (streq) at (domstreq1.north) {\bf Strong form (Eq. \ref{eq:dd-gov})};
\draw[<-, line width=0.5,] (0 + 2.8*\compsize, -1 - 2.1*\compsize) -- ++(1.0, -0.5)
    node[anchor=west,] (ifstreq) {$\jump{\tbq} = \avg{\bn\cdot\nabla\tbq} = 0$};
\node[anchor=south,] at (ifstreq.north) {Interface (Eq. \ref{eq:dd-gov}b--c)};

\draw[->, double, line width=0.5,] (streq.east) -- ++(0.5, 0) node[anchor=west,] {\bf Weak form (Eq. \ref{eq:weak-gov})};
\node[anchor=west, xshift=1.5cm,] at (domstreq.east) {$B_m$, $L_m$ in (Eq. \ref{eq:dd-gov-op})};
\node[anchor=west, xshift=0.8cm,] at (ifstreq.east) {$B_{m,n}$ in (Eq. \ref{eq:dd-gov-op})};

\end{tikzpicture}
    \caption{Illustration of domain decomposition.}
    \label{fig:dd-illustration}
\end{figure}
We consider the global-scale domain $\Omega$ decomposed into $M$ subdomains $\Omega_m$,
\begin{equation}
    \Omega = \bigcup\limits_{m=1}^M\Omega_m,
\end{equation}
where all subdomains can be categorized into a few reference domains,
\begin{equation}\label{eq:ref-domain}
    \Omega_m \in \rC\equiv\left\{\bOmega_1, \bOmega_2, \ldots\right\}
    \qquad \forall m=1,\ldots,M.
\end{equation}
The physics system state $\tbq$ is composed of subdomain state $\tbq_m \in H_1(\Omega_m)$,
\begin{equation}
    \tbq = \{\tbq_m\}_{m=1}^M.
\end{equation}
Each $\tbq_m$ satisfies the governing equation on its subdomain,
\begin{subequations}\label{eq:dd-gov}
    \begin{equation}
        \cL[\tbq_m] = \f_m \qquad \text{in } \Omega_m,
    \end{equation}
    with the continuity and smoothness constraints on the interface $\Gamma_{m,n}\equiv\partial\Omega_m\cap\partial\Omega_n$,
    \begin{equation}
        \jump{\tbq} \equiv \tbq_m - \tbq_n = 0 \qquad \text{on } \Gamma_{m,n}
    \end{equation}
    \begin{equation}
        \avg{\bn\cdot\nabla\tbq} \equiv \frac{1}{2}\left( \bn_m\cdot\nabla\tbq_m + \bn_n\cdot\nabla\tbq_n \right) = 0 \qquad \text{on } \Gamma_{m,n},
    \end{equation}
\end{subequations}
where $\bn_m$ is the outward normal vector of the subdomain $\Omega_m$,
and $\bn_m = -\bn_n$ on $\Gamma_{m,n}$.
Figure~\ref{fig:dd-illustration} illustrates (\ref{eq:dd-gov})
with an example of 3-by-3 global domain constructed with 3 types of components.

\subsection{Discontinuous Galerkin (DG) domain decomposition}\label{subsec:dg-dd}
Standard Galerkin finite-element discretization seeks an approximate solution $\bq\in\rQ_s$
that satisfies (\ref{eq:gov}) in a weak sense,
\begin{equation}\label{eq:weak-gov}
    \form{B}{\bqs}{\cL[\bq]} = \form{L}{\bqs}{\f},
\end{equation}
for all test function $\bqs\in\rQ_s$.
The Dirichlet boundary condition will be weakly enforced via DG formulation.
$\rQ_s \subset H^1(\Omega)$ is the piecewise $s$-th order polynomial function space
composed of $M$ subdomain finite element spaces,
\begin{equation}\label{eq:fes}
    \rQ_s = \prod_{m=1}^M \rQ_{m, s}.
\end{equation}
The specific definition of $\rQ_{m, s}$ will be introduced in the applications in Section~\ref{sec:poisson}~and~\ref{sec:stokes}.
The bilinear form $\form{B}{\cdot}{\cdot}$
and the linear form $\form{L}{\cdot}{\cdot}$
are the global-scale weak forms corresponding to the operators $\cL$, $\cN$ and the right-hand side $\f$, respectively.
\par
We seek a weak form (\ref{eq:weak-gov}) that approximates (\ref{eq:dd-gov}) in place of (\ref{eq:gov}),
where each global-scale weak form is decomposed for subdomains and their interfaces.
First, the bilinear form $\form{B}{\cdot}{\cdot}$ is decomposed as
\begin{subequations}\label{eq:dd-gov-op}
    \begin{equation}
    \begin{split}
        \form{B}{\bqs}{\cL[\bq]} =&
        \sum\limits_{m=1}^{M}\form{B_m}{\bqs}{\cL[\bq]} + 
        \sum\limits_{(m,n)\in \rI}\form{B_{m,n}}{\bqs}{\cL[\bq]}\\
        &+ \sum\limits_{m=1}^{M}\form{B_{m,di}}{\bqs}{\bq},
    \end{split}
    \end{equation}
where $B_m$ is the subdomain bilinear form defined on $\Omega_m$,
$B_{m,n}$ is the interface bilinear form defined on all interfaces $\Gamma_{m,n}$ with $\rI=\left\{(m,n) \,\middle|\, \Gamma_{m,n}\ne\varnothing\right\}$,
and $B_{m,di}$ is the Dirichlet boundary bilinear form defined on $\Gamma_{m,di} \equiv \partial\Omega_m \cap \partial\Omega_{di}$.
\par
The linear form $\form{L}{\cdot}{\cdot}$ is decomposed into subdomain linear forms $L_m$
and the boundary linear forms,
    \begin{equation}
        \form{L}{\bqs}{\f} = 
        \sum\limits_{m=1}^{M} \form{L_m}{\bqs}{\f} +
        \form{L_{ne}}{\bqs}{\f} + \form{L_{di}}{\bqs}{\f},
    \end{equation}
where the Neumann boundary linear form $L_{ne}$ is defined on $\partial\Omega_{ne}$,
and the Dirichlet boundary linear form $L_{di}$ on $\partial\Omega_{di}$.
\end{subequations}
The specific definitions of these operators will be introduced in the applications in Section~\ref{sec:poisson} and \ref{sec:stokes}.
Figure~\ref{fig:dd-illustration} illustrates
how these weak-form operators in (\ref{eq:dd-gov-op}) correspond to the strong-form equation (\ref{eq:dd-gov}).
\par
In DG domain decomposition~\cite{Arnold1982,Toselli2002,Hansbo2005}, the interface operators in (\ref{eq:dd-gov-op})
act as a penalty against the violation of the interface constraints (\ref{eq:dd-gov}b--c),
restricting the violation below discretization error.
Thus, it enforces the interface constraints (\ref{eq:dd-gov}b--c) only weakly,
without introducing Lagrange multiplier variables for the constraints
as in other strongly enforced domain decomposition methods~\cite{Farhat1991, Farhat1995}.
This can be applied flexibly to general Galerkin methods:
the subdomain operators $B_m$
and $L_m$ can be defined both in a continuous and in a discontinuous manner.
If extending an existing finite-element solver,
only the interface and boundary operators need to be implemented.
Furthermore, such weak enforcement can also bring an advantage in model reduction.
The reduced model significantly decreases the number of unknowns in the system,
making the strong interface enforcement a stringent constraint for such an overdetermined system,
worsening the overall accuracy~\cite{Hoang2021,diaz2023fast}.
Hoang~\textit{et al.}~\cite{Hoang2021} reported that
weakly constraining the interface condition improves the accuracy of the reduced model.
While their weak enforcement is based on a random sampling, somewhat far from the physics,
DG domain decomposition introduces the weak enforcement from the baseline governing equation,
ensuring the violation remains within the discretization error.
\par
These weak forms can also be written as matrix-vector inner products,
which explicitly show the operators that will be projected onto the reduced linear subspace in the subsequent section.
For bilinear operators,
\begin{subequations}\label{eq:comp-bilinear-mat}
    \begin{equation}
        \form{B_m}{\bqs}{\cL[\bq]} = \bq_m^{\dagger\top}\bB_m\bq_m
    \end{equation}
    \begin{equation}
        \form{B_{m,n}}{\bqs}{\cL[\bq]} = 
        \begin{pmatrix}
            \bq_m^{\dagger\top} & \bq_n^{\dagger\top}
        \end{pmatrix}
        \begin{pmatrix}
            \bB_{mm} & \bB_{mn} \\
            \bB_{nm} & \bB_{nn} \\
        \end{pmatrix}
        \begin{pmatrix}
            \bq_m \\ \bq_n
        \end{pmatrix}
    \end{equation}
    \begin{equation}
        \form{B_{m,di}}{\bqs}{\cL[\bq]} = \bq_m^{\dagger\top}\bB_{m,di}\bq_m,
    \end{equation}
\end{subequations}
where $\bB_m, \bB_{m,di} \in \rR^{N_m\times N_m}$ with $N_m\equiv\dim(\bq_m)$ degrees of freedom for $\bq_m$,
and $\bB_{i,j} \in \rR^{N_i\times N_j}$.
For linear form operators,
\begin{subequations}\label{eq:comp-linear-vec}
    \begin{equation}
        \form{L_m}{\bqs}{\f} = \bq_m^{\dagger\top}\bL_m[\f]
    \end{equation}
    \begin{equation}
        \form{L_{ne}}{\bqs}{\f} = \bq^{\dagger\top}\bL_{ne}[\f]
    \end{equation}
    \begin{equation}
        \form{L_{di}}{\bqs}{\f} = \bq^{\dagger\top}\bL_{di}[\f],
    \end{equation}
\end{subequations}
where $\bL_m: L_2(\Omega_m)\to\rR^{N_m}$ and $\bL_{ne}, \bL_{di}: L_2(\Omega)\to\rR^{N}$
with $N\equiv\dim{\bq}$ degrees of freedom for the global solution $\bq$.
\par
The weak-form operators in (\ref{eq:comp-bilinear-mat}--\ref{eq:comp-linear-vec})
lay the groundwork for constructing reduced-order model components,
which then can be assembled into a global scale reduced-order model.
The construction and assembly procedure of the reduced-order model
is described subsequently.

\subsection{Linear subspace model reduction}\label{eq:ls-rom}
\begin{figure}[tbph]
    \input{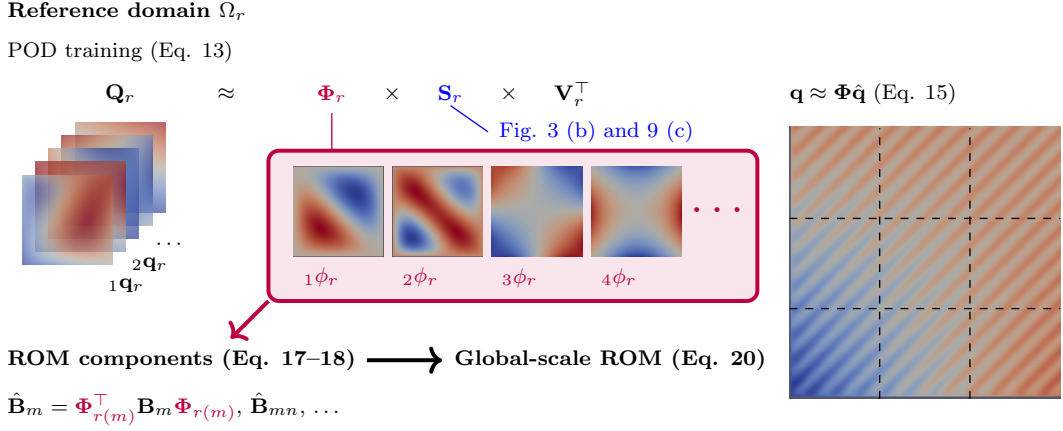}
    \caption{Diagram of the component reduced-order modeling procedure.}
    \label{fig:crom-diagram}
\end{figure}
Since all subdomains can be categorized into a few reference domains (\ref{eq:ref-domain}),
we attempt to identify a linear subspace for each reference domain $\bOmega_r$,
the basis for the reduced model component.
All sample solutions for the decomposed system (\ref{eq:dd-gov}) are first
categorized by their reference domains,
constituting the corresponding snapshot matrix $\bQ_r\in\rR^{N_r\times S_r}$,
\begin{subequations}\label{eq:snapshots}
    \begin{equation}
        \bQ_r =
        \left(\begin{array}{c:c:c}
            & & \\
            {}_1\bq_r & {}_2\bq_r & \cdots\\
            & & \\
        \end{array}\right),
    \end{equation}
    where all sample solutions ${}_i\bq_r$ lie on the reference domain $\bOmega_r$, i.e.,
    \begin{equation}
        \mathrm{dom}({}_i\bq_r) = \bOmega_r \qquad \forall i.
    \end{equation}
\end{subequations}
Specific sampling procedures for generating $\bQ_r$ will be introduced
in each application in Section~\ref{sec:poisson}~and~\ref{sec:stokes}.
The well-known proper orthogonal decomposition (POD) is used to decompose $\bQ_r$~\cite{Chatterjee2000,Liang2002,Rowley2017},
\begin{equation}\label{eq:pod}
\begin{split}
    \bQ_r &= \tilde{\bPhi}_r\tilde{\bS}_r\tilde{\bV}_r^{\top}\\
    &=
    \left(\begin{array}{c:c:c}
        & & \\
        {}_1\bphi_r & {}_2\bphi_r & \cdots\\
        & & \\
    \end{array}\right)
    \begin{pmatrix}
        {}_1\sigma_r & & \\
        & {}_2\sigma_r & \\
        & & \cdots \\
    \end{pmatrix}
    \left(\def\arraystretch{1.2}\begin{array}{ccc}
        & {}_1\bv_r^{\top} & \\\hdashline[2pt/2pt]
        & {}_2\bv_r^{\top} & \\\hdashline[2pt/2pt]
        & \vdots & \\
    \end{array}\right).
\end{split}
\end{equation}
The column vectors of $\tilde{\bPhi}_r$ correspond to the inherent physics modes we attempt to identify,
spanning the linear space on which all the snapshots ${}_i\bq_r$ lie.
The singular value ${}_i\sigma_r$ indicates the average magnitude of the basis vector ${}_i\bphi_r$ over all the snapshots in $\bQ_r$.
The right singular vector ${}_i\bv_r$ shows how the composition of ${}_i\bphi_r$ varies over each snapshot ${}_i\bq_r$,
which does not play a major role here.
\par
The singular values $\tilde{\bS}_r$ provide meaningful information from the snapshot data for constructing the linear subspace.
In essence, the singular value ${}_i\sigma_r$ corresponds to the total energy of the mode ${}_i\bphi_r$ within the snapshot sample $\bQ_r$~\cite{Rowley2017}.
In other words, the ${}_i\bphi_r$ with the larger ${}_i\sigma_r$ corresponds to the most dominant physics modes representing the snapshots.
In many applications, the singular values of a few major ${}_i\bphi_r$ can cover a significant portion of the norm of $\bQ_r$,
indicating that the snapshots in $\bQ_r$ can be well explained with only a few major modes.
We therefore attempt to approximate $\bQ_r$ in the linear subspace of a reduced number, say $R_r \ll S_r$, of basis vectors ${}_i\bphi_r$,
\begin{equation}\label{eq:pod-training}
\begin{split}
    \bQ_r &\approx \bPhi_r\bS_r\bV_r^{\top}\\
    &=
    \left(\begin{array}{c:c:c}
        & & \\
        {}_1\bphi_r & \cdots & {}_{R_r}\bphi_{r}\\
        & & \\
    \end{array}\right)
    \begin{pmatrix}
        {}_1\sigma_r & & \\
        & \cdots & \\
        & & {}_{R_r}\sigma_{r} \\
    \end{pmatrix}
    \left(\def\arraystretch{1.2}\begin{array}{ccc}
        & {}_1\bv_r^{\top} & \\\hdashline[2pt/2pt]
        & \vdots & \\\hdashline[2pt/2pt]
        & {}_{R_r}\bv_{r}^{\top} & \\
    \end{array}\right).
\end{split}
\end{equation}
Figure~\ref{fig:crom-diagram} illustrates this approximation with an example of the Poisson equation
that will be demonstrated in Section~\ref{sec:poisson}.
Any other solution we attempt to predict on $\bOmega_r$ is also approximated by this reduced number of basis vectors,
\begin{equation}
    \bq_r \approx \bPhi_r{\hbq}_r.
\end{equation}
The global-scale solution $\bq$ is approximated in each subdomain,
\begin{equation}\label{eq:global-rom-sol}
    \bq \approx \bPhi{\hbq} \equiv
    \begin{pmatrix}
        \bPhi_{r(1)} & & & & \\
        & \cdots & & & \\
        & & \bPhi_{r(m)} & & \\
        & & & \cdots & \\
        & & & & \bPhi_{r(M)} \\
    \end{pmatrix}
    \left(\begin{array}{c}
        \hbq_1 \\
        \vdots \\
        \hbq_m \\
        \vdots \\
        \hbq_M \\
    \end{array}\right),
\end{equation}
where $r(m)$ is the reference domain corresponding to the subdomain $m$.
\par
We project the decomposed system (\ref{eq:dd-gov}) onto the linear subspaces of $\bPhi$.
In particular, we consider Galerkin projection where the test function is also approximated with the same linear subspace,
\begin{equation}
    \bqs \approx \bPhi\hbqs.
\end{equation}
The matrix operators of the bilinear form in (\ref{eq:comp-bilinear-mat}) are then reduced to
\begin{subequations}\label{eq:comp-bilinear-mat-reduced}
    \begin{equation}\label{eq:comp-bilinear-mat-reduced:Bm}
    \begin{split}
        \form{B_m}{\bqs}{\cL[\bq]}
        &\approx \form{\hB_m}{\hbqs}{\hbq}\\
        &\equiv \hbq_m^{\dagger\top}\hbB_m\hbq_m
        \equiv \hbq_m^{\dagger\top}\left(\bPhi_{r(m)}^{\top}\bB_m\bPhi_{r(m)}\right)\hbq_m
    \end{split}
    \end{equation}
    \begin{equation}\label{eq:comp-bilinear-mat-reduced:Bmn}
    \begin{split}
        \form{B_{m,n}}{\bqs}{\cL[\bq]}
        &\approx \form{\hB_{m,n}}{\hbqs}{\hbq}\\
        &\equiv 
        \begin{pmatrix}
            \hbq_m^{\dagger\top} & \hbq_n^{\dagger\top}
        \end{pmatrix}
        \begin{pmatrix}
            \hbB_{mm} & \hbB_{mn} \\
            \hbB_{nm} & \hbB_{nn} \\
        \end{pmatrix}
        \begin{pmatrix}
            \hbq_m \\ \hbq_n
        \end{pmatrix}\\
        &\equiv 
        \begin{pmatrix}
            \hbq_m^{\dagger\top} & \hbq_n^{\dagger\top}
        \end{pmatrix}
        \begin{pmatrix}
            \bPhi_{r(m)}^{\top}\bB_{mm}\bPhi_{r(m)} & \bPhi_{r(m)}^{\top}\bB_{mn}\bPhi_{r(n)} \\
            \bPhi_{r(n)}^{\top}\bB_{nm}\bPhi_{r(m)} & \bPhi_{r(n)}^{\top}\bB_{nn}\bPhi_{r(n)} \\
        \end{pmatrix}
        \begin{pmatrix}
            \hbq_m \\ \hbq_n
        \end{pmatrix}
    \end{split}
    \end{equation}
    \begin{equation}
        \begin{split}
            \form{B_{m,d}}{\bqs}{\cL[\bq]}
            &\approx \form{\hB_{m,d}}{\hbqs}{\hbq}\\
            &\equiv \hbq_m^{\dagger\top}\hbB_{m,d}\hbq_m
            \equiv \hbq_m^{\dagger\top}\left(\bPhi_{r(m)}^{\top}\bB_{m,d}\bPhi_{r(m)}\right)\hbq_m.
        \end{split}
        \end{equation}
\end{subequations}
Likewise, the vectors of the linear form in (\ref{eq:comp-linear-vec}) are also reduced to
\begin{subequations}\label{eq:comp-linear-vec-reduced}
    \begin{equation}
    \begin{split}
        \form{L_m}{\bqs}{\f}
        &\approx \form{\hL_m}{\hbqs}{\f}\\
        &\equiv \hbq_m^{\dagger\top}\hbL_m[\f]
        \equiv \hbq_m^{\dagger\top}\left(\bPhi_{r(m)}^{\top}\bL_m[\f]\right)
    \end{split}
    \end{equation}
    \begin{equation}
    \begin{split}
        \form{L_n}{\bqs}{\f}
        &\approx \form{\hL_n}{\hbqs}{\f}\\
        &\equiv \hbq^{\dagger\top}\hbL_n[\f]
        \equiv \hbq^{\dagger\top}\left(\bPhi^{\top}\bL_n[\f]\right)
    \end{split}
    \end{equation}
    \begin{equation}
    \begin{split}
        \form{L_d}{\bqs}{\f}
        &\approx \form{\hL_d}{\hbqs}{\f}\\
        &\equiv \hbq^{\dagger\top}\hbL_d[\f]
        \equiv \hbq^{\dagger\top}\left(\bPhi^{\top}\bL_d[\f]\right).
    \end{split}
    \end{equation}
\end{subequations}
\par
(\ref{eq:comp-bilinear-mat-reduced}--\ref{eq:comp-linear-vec-reduced})
are the ROM components, which are assembled into
the global-scale reduced-model operators analogous to (\ref{eq:dd-gov-op}),
\begin{subequations}\label{eq:dd-reduced-op}
    \begin{equation}
        \form{\hB}{\hbqs}{\hbq} =
        \sum\limits_{m=1}^{M}\form{\hB_m}{\hbqs}{\hbq} + 
        \sum\limits_{(m,n)\in \rI}\form{\hB_{m,n}}{\hbqs}{\hbq} +
        \sum\limits_{m=1}^{M}\form{\hB_{m,d}}{\hbqs}{\hbq}
    \end{equation}
    \begin{equation}
        \form{\hL}{\hbqs}{\f} = 
        \sum\limits_{m=1}^{M} \form{\hL_m}{\hbqs}{\f} +
        \form{\hL_n}{\hbqs}{\f} + \form{\hL_d}{\hbqs}{\f}.
    \end{equation}
\end{subequations}
The global-scale reduced model analogous to (\ref{eq:weak-gov}) is
\begin{equation}\label{eq:weak-gov-reduced}
    \form{\hB}{\hbqs}{\hbq} = \form{\hL}{\hbqs}{\f},
\end{equation}
which is solved for $\hbq$ with respect to all $\hbqs$.
Figure~\ref{fig:crom-diagram} illustrates the procedure of global-scale ROM construction.
\par
It is worth emphasizing that the global-scale full-order model (\ref{eq:weak-gov}) need not be solved at all
for constructing the reduced model (\ref{eq:weak-gov-reduced}).
Neither does it need to be constructed at all, throughout the entire reduced model construction.
The entire procedure can be executed only at the reference component domain level.
The snapshots in $\bQ_r$ (\ref{eq:snapshots}) can be obtained by
solving (\ref{eq:weak-gov}) only on $\bOmega_r$ with appropriate $\f$ and boundary conditions.
The reduced basis $\bPhi_r$ is then calculated only from this component-scale snapshots.
Any subdomain can be categorized with a reference domain per (\ref{eq:ref-domain}),
so it is possible to pre-compute the reduced operators in (\ref{eq:dd-reduced-op})
with respect to the reference domains and their interfaces.
For example, $\form{\hB_m}{\hbqs}{\hbq}$ in (\ref{eq:comp-bilinear-mat-reduced:Bm})
can use the pre-computed operator $\hbB_{r(m)}$,
\begin{equation}\label{eq:rom-precompute}
\begin{split}
    \form{\hB_m}{\hbqs}{\hbq}
    &= \hbq_m^{\dagger\top}\hbB_{r(m)}\hbq_m
    \equiv \hbq_m^{\dagger\top}\left(\bPhi_{r(m)}^{\top}\bB_{r(m)}\bPhi_{r(m)}\right)\hbq_m,
\end{split}
\end{equation}
where $\bB_{r(m)}$ can be constructed with the reference domain $\bOmega_r$ only.
Likewise, the interface operator $\hB_{r,r'}$ in (\ref{eq:comp-bilinear-mat-reduced:Bmn})
can be precomputed for an interface between $\bOmega_r$ and $\bOmega_{r'}$,
and used for all $(m, n)$ with $r(m) = r$ and $r(n) = r'$.
\section{Poisson equation}\label{sec:poisson}
\subsection{Governing equation}
For $\tbq\equiv \tilde{u}\in H^1(\Omega)$, $\f\equiv f \in L_2(\Omega)$,
the governing equation (\ref{eq:gov}) for the Poisson equation is
\begin{subequations}\label{eq:poisson-gov}
    \begin{equation}
        -\nabla^2 \tilde{u} = f \qquad \text{in } \Omega,
    \end{equation}
    with boundary conditions on $\partial\Omega = \partial\Omega_{di} \cup \partial\Omega_{ne}$,
    \begin{equation}
        \tilde{u} = g_{di} \qquad \text{on } \partial\Omega_{di},
    \end{equation}
    \begin{equation}
        \bn\cdot\nabla\tilde{u} = g_{ne} \qquad \text{on } \partial\Omega_{ne}.
    \end{equation}
\end{subequations}
(\ref{eq:poisson-gov}) is decomposed into subdomains as in (\ref{eq:dd-gov}),
\begin{subequations}\label{eq:poisson-dd-gov}
    \begin{equation}
        -\nabla^2 \tilde{u}_m = f \qquad \text{in } \Omega_m
    \end{equation}
    with interface constraints on $\Gamma_{m,n}$,
    \begin{equation}
        \jump{\tilde{u}} = 0,
    \end{equation}
    \begin{equation}
        \avg{\bn\cdot\nabla\tilde{u}} = 0.
    \end{equation}
\end{subequations}

\subsection{Weak form finite element formulation}
For the interface handling we follow the formulation of Hansbo~\cite{Hansbo2005},
though other DG-type domain decomposition techniques can be used as well.
Our choice of the finite element space (\ref{eq:fes}) is
\begin{equation}\label{eq:poisson-fes}
    \rQ_{m, s} = \left\{ u_m \in H^1(\Omega_m) \;\bigg|\; u_m\big|_{\kappa} \in V_s(\kappa) \quad \forall \kappa\in\cT(\Omega_m) \right\},
\end{equation}
where $\cT(\Omega_m)$ is the set of mesh elements in subdomain $\Omega_m$,
and $V_s(\kappa)$ is the space of $s$th-order polynomials in a mesh element $\kappa$.
Under this choice, the FE formulation of each subdomain remains continuous Galerkin,
enforcing only the interface constraints in a discontinuous Galerkin way.
However, another choice is also available where the entire system is formulated as a DG FE discretization.
\par
The weak form of (\ref{eq:poisson-dd-gov}) corresponding to (\ref{eq:weak-gov})
is formulated with respect to $\bqs=\{\us_m\}$ and $\bq=\{u_m\}$.
The weak-form operators in (\ref{eq:dd-gov-op}) are defined as
\begin{subequations}\label{eq:poisson-B-comp}
    \begin{equation}
        \form{B_m}{\bqs}{\cL[\bq]} = \inprod{\nabla \us_m}{\nabla u_m}_{\Omega_m},
    \end{equation}
    \begin{equation}
    \begin{split}
        \form{B_{m,n}}{\bqs}{\cL[\bq]} =&
        - \inprod{\avg{\bn\cdot\nabla \us}}{\jump{u}}_{\Gamma_{m,n}}
        - \inprod{\jump{\us}}{\avg{\bn\cdot\nabla u}}_{\Gamma_{m,n}}\\
        &+ \inprod{\gamma\Delta\bx^{-1}\jump{\us}}{\jump{u}}_{\Gamma_{m,n}},
    \end{split}
    \end{equation}
    with $\Delta\bx$ the diameter of the elements and $\gamma > 0$ the penalty strength for interface constraints.
    \todo[inline]{PR: Is diameter commonly used for $\Delta x$? If not, then length or size of the elements can be an option. Both are commonly used in FV literature.\\ KC: Diameter here refers to a mathematical term in FE community, which lead to the length of the element.}
    On the Dirichlet boundary of the subdomain $\Gamma_{m,{di}} = \partial\Omega_m \cap \partial\Omega_{di}$,
    \begin{equation}
\begin{split}
        \form{B_{m,di}}{\bqs}{\bq} =&
        - \inprod{\bn\cdot\nabla \us}{u}_{\Gamma_{m,di}}
        - \inprod{\us}{\bn\cdot\nabla u}_{\Gamma_{m,di}}\\
        &+ \inprod{\gamma\Delta\bx^{-1}\us}{u}_{\Gamma_{m,di}}.
    \end{split}
    \end{equation}
\end{subequations}
The linear weak-form operators are defined as
\begin{subequations}\label{eq:poisson-L-comp}
    \begin{equation}
        \form{L_m}{\bqs}{\f} = \inprod{\us_m}{f}_{\Omega_m},
    \end{equation}
    \begin{equation}
    \begin{split}
        \form{L_{ne}}{\bqs}{\f} =&
        \inprod{\us}{g_{ne}}_{\partial\Omega_{ne}},
    \end{split}
    \end{equation}
    \begin{equation}
        \begin{split}
            \form{L_{di}}{\bqs}{\f} &=
            \inprod{\us}{\gamma\Delta\bx^{-1}g_{di}}_{\partial\Omega_{di}}
            + \inprod{\bn\cdot\nabla\us}{g_{di}}_{\partial\Omega_{di}}.
        \end{split}
        \end{equation}
\end{subequations}
The resulting weak form discretized system is positive definite for a sufficiently large $\gamma$~\cite{Hansbo2005},
where $\gamma = (s+1)^2$ is used for $\bq \equiv u \in \bQ_s$ in our application~\cite{mfem}.
The full order model (FOM) discretization (\ref{eq:dd-gov}) is implemented for (\ref{eq:poisson-dd-gov}) in the \texttt{MFEM} framework,
which is an open-source C++ library for finite element methods~\cite{mfem}.

\subsection{Component-level samples for linear subspace training}\label{subsec:poisson-training}
In this demonstration, we consider a single component of a two-dimensional unit square domain
$\rC=\{\bOmega_1\}$, $\bOmega_1 = [0, 1]^2$.
The domain $\bOmega_1$ is discretized into a mesh with $64^2$ uniform rectangular elements.
We consider $s = 1$, i.e. first-order polynomial function space for $\rQ_{m,s}$ (\ref{eq:poisson-fes}),
leading to $N_1 = 4225$ degrees of freedom for the FOM in $\bOmega_1$.
\par
For sample generation, the right-hand side is defined in a general sinusoidal form with parameters $\bk\in\rR^2$ and $\theta\in\rR$,
\begin{subequations}\label{eq:poisson-component}
    \begin{equation}
        f = \sin 2\pi(\bk\cdot\bx + \theta).
    \end{equation}
    The entire boundary is set as a Dirichlet condition, i.e. $\partial\bOmega_1 \equiv \partial\bOmega_{1, di}$,
    with parameters $\bk_b\in\rR^2$ and $\theta_b\in\rR$,
    \begin{equation}
        g_{di} = \sin 2\pi(\bk_b\cdot\bx + \theta_b).
    \end{equation}
\end{subequations}
The parameter values of $\bk$, $\bk_b$, $\theta$ and $\theta_b$
are randomly chosen for each sample,
thereby collecting a broad spectrum of solutions.
\begin{subequations}\label{eq:poisson-training-data}
    $\bk$ and $\bk_b$ are chosen from a uniform random distribution of two-dimensional vectors,
    \begin{equation}
        \bk, \bk_b \in U[-0.5, 0.5]^2,
    \end{equation}
    and $\theta$ and $\theta_b$ are \PR{chosen} from a scalar uniform random distribution,
    \begin{equation}
        \theta, \theta_b \in U[0, 1].
    \end{equation}
\end{subequations}
\par
\begin{figure}[tbh]
    \begin{tikzpicture}[font=\small,]
    \begin{groupplot}[
        group style={
            group name = my plots,
            group size= 2 by 1,
            xlabels at =edge bottom,
            horizontal sep=3.5cm,
            vertical sep=2.2cm,
        },
        name=chung,
    ]    
\pgfplotsset{set layers=standard}%

        \nextgroupplot[
            height = 0.45\textwidth,
            width = 0.45\textwidth,
            ylabel={$x_2$},
            xlabel={$x_1$},
            tick scale binop ={\times},
            xmin = 0, xmax = 1,
            ymin = 0, ymax = 1,
            point meta min=-1.0, point meta max=1.0,
            colorbar, colormap/viridis,
            colorbar style={
                font=\scriptsize,
                xticklabel pos=upper,
                scaled y ticks=false,
                /pgf/number format/precision=4,
                at={(rel axis cs: 1.01, 0.)}, anchor=south west,
                xlabel=$u(\bx)$,
            }
        ]
        
            \edef\imagepath{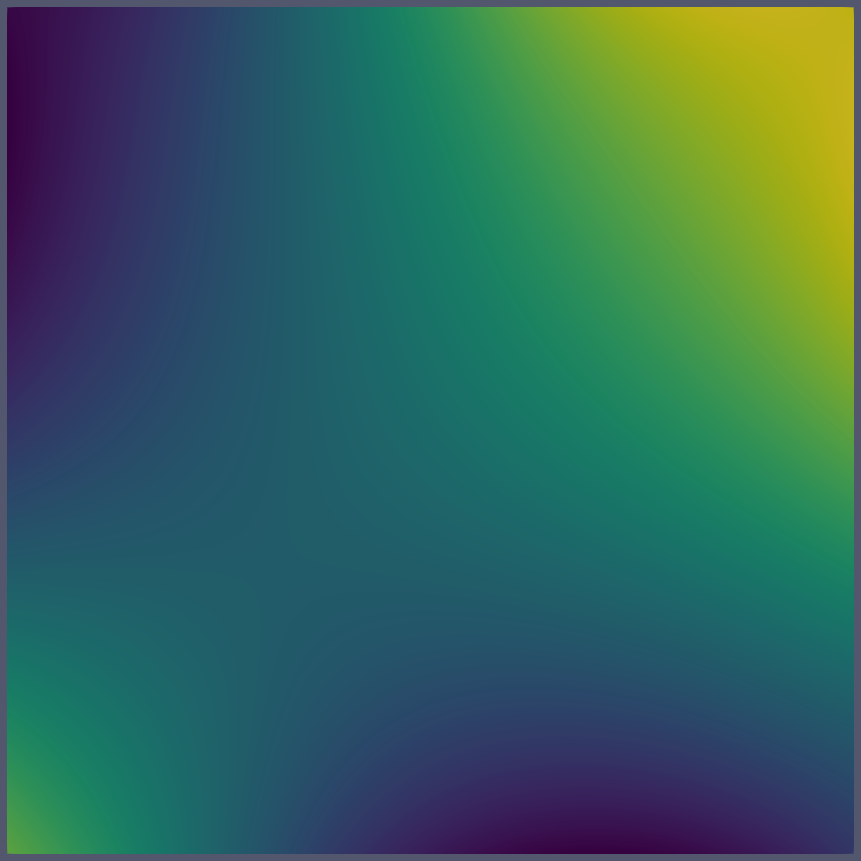}
            \addplot graphics[xmin=-0.01,xmax=1.01,ymin=-0.01,ymax=1.01]{\imagepath};

        \nextgroupplot[
            height = 0.45\textwidth,
            width = 0.5\textwidth,
            ylabel={Singular value ${}_i\sigma_1$},
            xlabel={POD mode $i$},
            tick scale binop ={\times},
            ymode=log,
            xmin=0, xmax=40,
            legend style={
                font=\small,
                draw=none, fill=none,
                at={(rel axis cs: 0., 1.0)},
                anchor=south west,
                nodes={scale=1.0},
                legend cell align={left},
                legend columns=4,
                /tikz/every even column/.append style={column sep=0.5cm},
            },
            legend image post style={mark options={scale=1.0, fill=white, line width=1.0}},
        ]
        
            \addplot+ [
                line width=1.0,
                smooth, solid,
                mark=*,
                mark options={fill=white,},
            ]
            table [x expr=\coordindex+1, y index=0]{./poisson_component_square_sv.txt};

            \draw[
                line width=0.5,
                gray, solid,
            ] (axis cs: 15, 1e4) -- node[anchor=west, black, font=\scriptsize] {$\Leftarrow$ $99.77\%$ coverage} (axis cs: 15, 1e0);
 
  \end{groupplot}
 \node[below = 1.5cm of my plots c1r1.south west,
            anchor=west,
        ] {(a) Sample solution};
\node[below = 1.5cm of my plots c2r1.south west,
            anchor=west,
        ] {(b) POD mode spectrum};
\end{tikzpicture}
%
    \caption{Sample generation and linear subspace training for the Poisson equation:
    (a) a sample solution in the component domain $\bOmega_1$ with $\bk = (0.4, -0.3)$, $\bk_b = (0.5, 0.4)$, $\theta=0.1$ and $\theta_b=0.4$;
    and (b) Singular value spectrum of POD basis identified from $\bQ_1$.
    The gray line indicates the $15$-th singular value, up to which $99.77\%$ of spectrum is covered.}
    \label{fig:poisson-sample}
\end{figure}
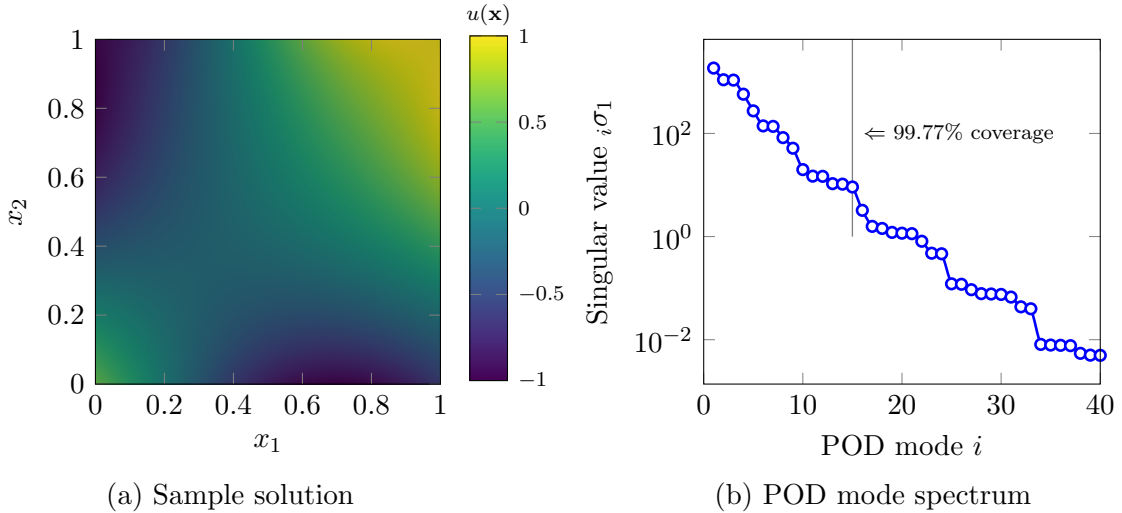
Figure~\ref{fig:poisson-sample}(a) shows a sample solution with randomly chosen parameter values.
In total, $S_1 = 4225$ samples are collected for the snapshot matrix $\bQ_1$.
From $\bQ_1$, the linear subspace basis vectors are identified via POD (\ref{eq:pod}),
under the \texttt{libROM} framework, which is an open-source library for ROM~\cite{librom}.
Figure~\ref{fig:poisson-sample}(b) shows the singular values $\mathrm{diag}(\tilde{\bS}_1)$
associated with the identified POD basis $\tilde{\bPhi}_1$.
The singular value spectrum decays rapidly,
and $99.77\%$ of the energy norm of the snapshot matrix is covered by only the first 15 basis vectors, i.e.
\begin{equation}
    \sum\limits_{i=1}^{15} {}_i\sigma_1 \approx 0.9977\sum\limits_{s=1}^{S_1}\Vert {}_s\bq_1\Vert.
\end{equation}
From this we \PR{can} expect that any test case can be effectively spanned by these first 15 basis vectors,
unless it is significantly different from these snapshots.
In practice, even the qualitatively different case can be robustly predicted with the component ROM constructed from this snapshot.
This will be shown in the subsequent demonstration.

\subsection{Prediction performance}
The ROM component (\ref{eq:comp-bilinear-mat-reduced}--\ref{eq:comp-linear-vec-reduced})
for the component $\bOmega_1$ is built by
projecting the component operator (\ref{eq:poisson-B-comp}) and (\ref{eq:poisson-L-comp})
onto the basis $\bPhi_1$, which is obtained as described in Section~\ref{subsec:poisson-training}.
With the resulting ROM component, the global ROM (\ref{eq:weak-gov-reduced}) is then constructed per (\ref{eq:dd-reduced-op})
at various sizes of the global domain,
and tested for various right-hand sides $f$ of (\ref{eq:poisson-gov}).
\par
In the following computational experiments,
the comparison between FOM and ROM is made on an Intel Xeon CPU, running on a single thread with 3 GB of RAM and 2.1 GHz clock speed~\cite{quartz}.

\subsubsection{Global-level prediction}\label{subsubsec:poisson-scaleup}
\begin{figure}[tbh]
    \input{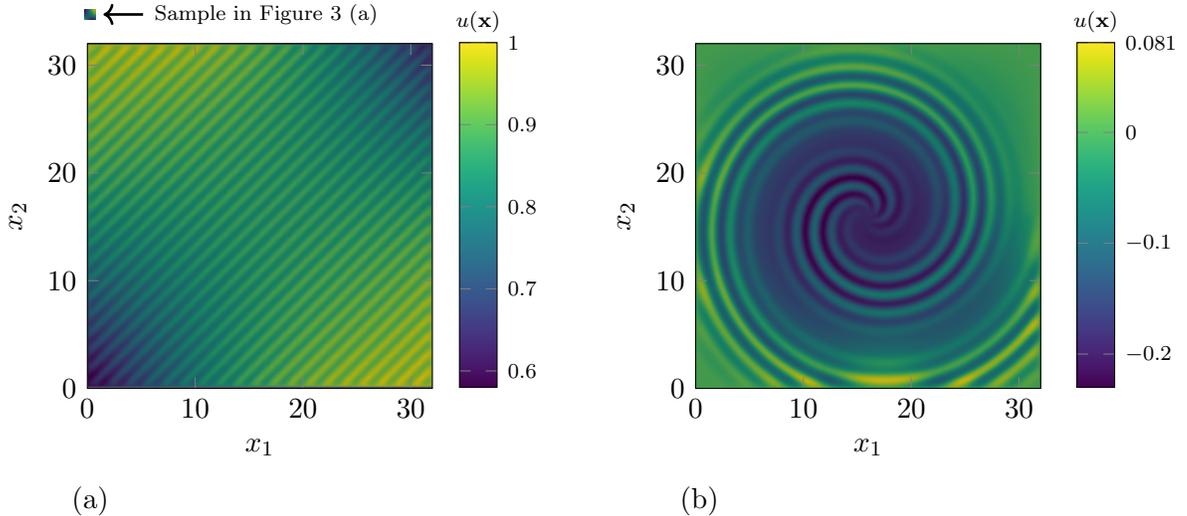}
    \caption{Example predictions on the global domain $\Omega=[0,32]^2$ with the global ROM (\ref{eq:weak-gov-reduced})
    constructed with $32^2$ components: ROM solutions for
    (a) the component problem (\ref{eq:poisson-component}) with $\bk=(0.6, -0.6)$, $\bk_b=(0.05, 0.04)$, $\theta=0.1$ and $\theta_b=0$;
    and for (b) the Spiral problem (\ref{eq:poisson-spiral}) with $(s, k) = (1.53, 3)$.
    The component-level sample from Figure~\ref{fig:poisson-sample} is included as a reference to compare size.
    \todo[inline]{PR: The subcaption label numbers (a,b) are not centered probably because you removed the subcaptions\\ KC: It's regardless of subcaptions. I adjusted them a little bit.}
    }
    \label{fig:poisson-global}
\end{figure}
The global ROM (\ref{eq:weak-gov-reduced}) for Poisson equation (\ref{eq:poisson-gov})
is first demonstrated for the component problem (\ref{eq:poisson-component}),
which was also used to generate the component-level samples.
Six different sizes of the global domain are considered:
$\Omega = [0, 2^n]^2 \subset \rR^2$ with $M = n^2$ components of $\bOmega_1$
for $n = 2, 3, \ldots, 7$.
For each size of the global domain,
in order to investigate the robustness of prediction outside the training range,
100 test values of the parameters $\{\bk, \bk_b, \theta, \theta_b\}$
are chosen from a range larger than the training data (\ref{eq:poisson-training-data}),
\begin{subequations}
    \begin{equation}
        \bk, \bk_b \in [-0.7, 0.7]^2
    \end{equation}
    \begin{equation}
        \theta, \theta_b \in [0, 1]^2.
    \end{equation}
\end{subequations}
Figure~\ref{fig:poisson-global}~(a) shows an example test prediction on the global domain of size $[0, 32]^2$.
\par
\begin{figure}[htbp]
    \begin{tikzpicture}[font=\small,]
    \begin{groupplot}[
        group style={
            group name = my plots,
            group size= 2 by 2,
            xlabels at =edge bottom,
            horizontal sep=3.5cm,
            vertical sep=2.2cm,
        },
        name=chung,
    ]    
\pgfplotsset{set layers=standard}%

        \nextgroupplot[
            height = 0.45\textwidth,
            width = 0.5\textwidth,
            xlabel={Number of components},
            ylabel={Assembly time ($s$)},
            tick scale binop ={\times},
            xmode=log, ymode=log,
            legend style={
                font=\small,
                draw=none, fill=none,
                at={(rel axis cs: 0., 1.0)},
                anchor=south west,
                nodes={scale=1.0},
                legend cell align={left},
                legend columns=4,
                /tikz/every even column/.append style={column sep=0.5cm},
            },
            legend image post style={mark options={scale=1.0, fill=white, line width=1.0}},
        ]
        


            \addplot+ [
                line width=1.0,
                solid,
                mark=*,
                mark options={fill=white,},
                blue,
                error bars/.cd, y dir=both, y explicit,
            ]
            table [
                x expr=\thisrowno{0}^2, y index=1,
                y error minus expr=\thisrowno{1} - \thisrowno{4},
                y error plus expr=\thisrowno{5} - \thisrowno{1},
            ]{./poisson_direct_rom_assemble.txt};

            \addplot+ [
                line width=1.0,
                solid,
                mark=*,
                mark options={fill=white,},
                red,
                error bars/.cd, y dir=both, y explicit,
            ]
            table [
                x expr=\thisrowno{0}^2, y index=1,
                y error minus expr=\thisrowno{1} - \thisrowno{4},
                y error plus expr=\thisrowno{5} - \thisrowno{1},
            ]{./poisson_direct_fom_assemble.txt};

            \legend{ROM, FOM}

        \nextgroupplot[
            height = 0.45\textwidth,
            width = 0.5\textwidth,
            xlabel={Number of components},
            ylabel={Solution time ($s$)},
            tick scale binop ={\times},
            xmode=log, ymode=log,
        ]
        


            \addplot+ [
                line width=1.0,
                solid,
                mark=*,
                mark options={fill=white,},
                blue,
                error bars/.cd, y dir=both, y explicit,
            ]
            table [
                x expr=\thisrowno{0}^2, y index=1,
                y error minus expr=\thisrowno{1} - \thisrowno{4},
                y error plus expr=\thisrowno{5} - \thisrowno{1},
            ]{./poisson_direct_rom_solve.txt};

            \addplot+ [
                line width=1.0,
                solid,
                mark=*,
                mark options={fill=white,},
                red,
                error bars/.cd, y dir=both, y explicit,
            ]
            table [
                x expr=\thisrowno{0}^2, y index=1,
                y error minus expr=\thisrowno{1} - \thisrowno{4},
                y error plus expr=\thisrowno{5} - \thisrowno{1},    
            ]{./poisson_direct_fom_solve.txt};

            \logLogSlopeTriangle{0.6}{0.1}{0.38}{1}{}{north}
 
        \nextgroupplot[
            height = 0.45\textwidth,
            width = 0.5\textwidth,
            xlabel={Number of components},
            ylabel={Relative error ($\%$)},
            tick scale binop ={\times},
            xmode=log, ymode=log,
            ymin=1e-2, ymax=1e1,
            xshift=4cm,
            legend pos=south west,
            legend style={
                font=\small,
                draw=none, fill=none,
                anchor=south west,
                nodes={scale=1.0},
                legend cell align={left},
            },
        ]

            \addlegendimage{
                line width=1.0,
                draw=none,
                mark=square*,
                mark options={fill=white,},
                brown,
            }
            \addlegendentry{Spiral problem (Eq. \ref{eq:poisson-spiral})}
        
            \addplot+ [
                line width=1.0,
                solid,
                mark=*,
                mark options={fill=white,},
                blue,
                error bars/.cd, y dir=both, y explicit,
            ]
            table [
                x expr=\thisrowno{0}^2, y expr=\thisrowno{1} * 1e2,
                y error minus expr=(\thisrowno{1} - \thisrowno{4}) * 1e2,
                y error plus expr=(\thisrowno{5} - \thisrowno{1}) * 1e2,
            ]{./poisson_direct_rel_error.txt};

            \addplot+ [
                line width=1.0,
                draw=none,
                mark=square*,
                mark options={fill=white,},
                brown,
                error bars/.cd, y dir=both, y explicit,
            ]
            table [
                x expr=\thisrowno{0}^2, y expr=\thisrowno{1} * 1e2,
                y error minus expr=(\thisrowno{1} - \thisrowno{4}) * 1e2,
                y error plus expr=(\thisrowno{5} - \thisrowno{1}) * 1e2,
            ]{./spiral.rel_error.txt};

  \end{groupplot}
\node[below = 1.5cm of my plots c1r1.south west,
    anchor=west,
] {(a)};
\node[below = 1.5cm of my plots c2r1.south west,
    anchor=west,
] {(b)};
\node[below = 1.5cm of my plots c1r2.south west,
    anchor=west, xshift=4cm,
] {(c)};
\end{tikzpicture}
%
    \caption{Performance of the ROM compared to the FOM for the Poisson equation, depending on the size of the domain:
    (a) assembly time of the system;
    (b) computation time of the system; and
    (c) relative error of the ROM compared to the FOM solution.
    The marker denotes the median value of 100 test cases,
    and the error bar denotes $95\%$-confidence interval of the test cases.
    Both ROM and FOM systems are solved using the conjugate-gradient method without any preconditioner.
    }
    \label{fig:poisson-scaleup}
\end{figure}
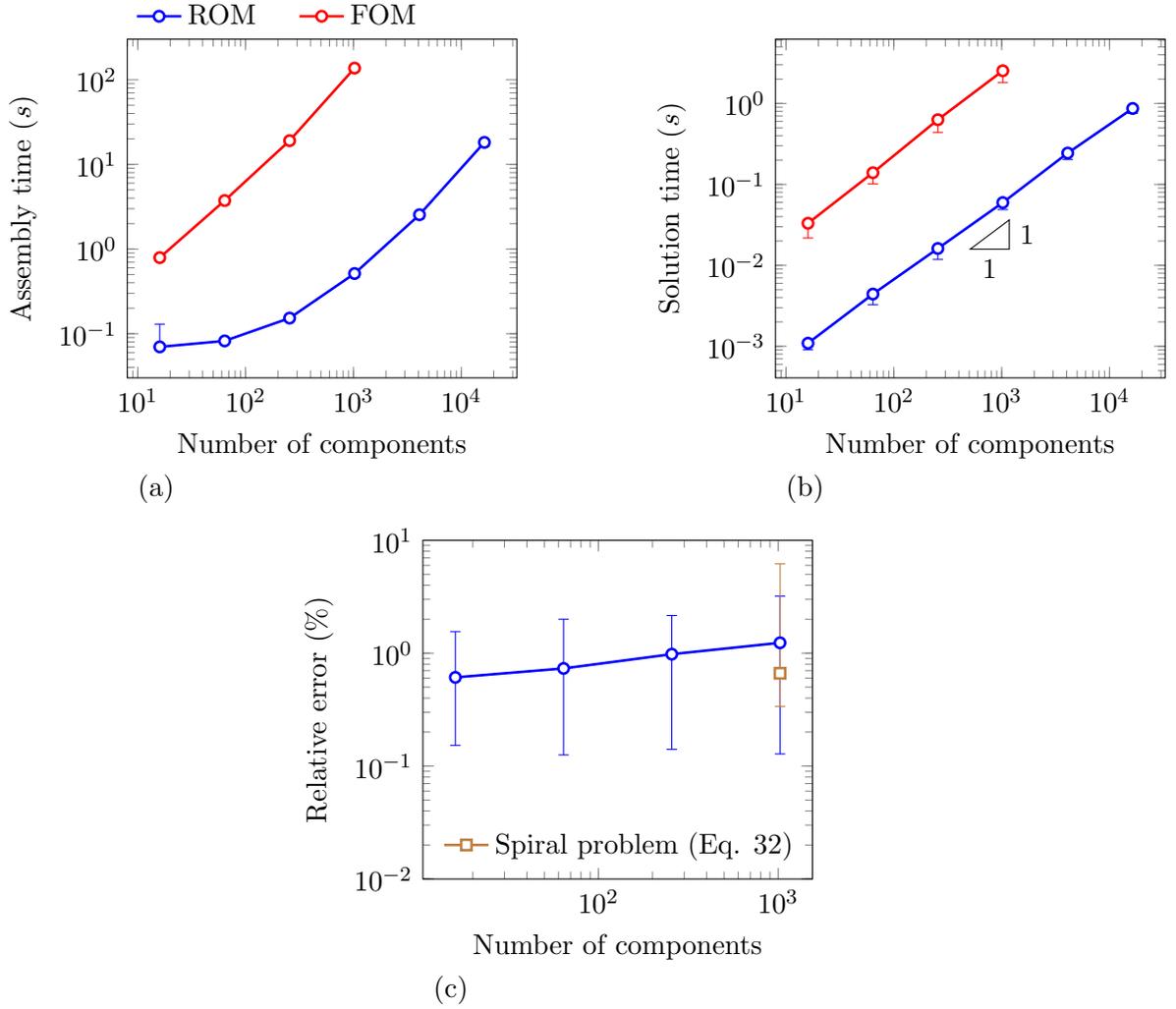
Figure~\ref{fig:poisson-scaleup} shows the comparison between the FOM and the global ROM,
depending on the size of the global domain.
Here, both systems are solved using \texttt{MUMPS}, which is a direct LU-factorization linear solver~\cite{Amestoy2001,Amestoy2019}.
In Figure~\ref{fig:poisson-scaleup}~(a),
assembly time for both FOM and ROM system scales super-linearly with the number of components.
The ROM global system does not need the assembly of the FOM global system,
since the ROM operators are pre-computed per each reference as described in Section~\ref{eq:ls-rom} (Eq. \ref{eq:rom-precompute}).
Furthermore, the size of ROM operators ($\rR^{15\times15}$) is significantly smaller than that of FOM operators ($\rR^{4225\times4225}$),
leading to a huge reduction of assembly time.
For $M>32^2$,
the FOM system could not be assembled due to the memory limitations.
On the other hand, the ROM system can be assembled for the larger domains under the same memory limitations,
which is confirmed up to an $M=128^2$-component domain.
Except for a few outliers, the assembly times of both FOM and ROM are consistent over all 100 test cases.
\par
Figure~\ref{fig:poisson-scaleup}~(b) compares the computation time of solving the ROM and FOM systems.
Compared to the FOM,
the ROM accelerates the computation time about 30 times faster at $M=16$.
The speed-up factor increases up to 40 with the number of components, up to $M=1024$.
This speed-up trend was consistent over all 100 test cases, without any outlier.
Meanwhile, the prediction accuracy of the ROM does not degrade regardless of the domain size.
Figure~\ref{fig:poisson-scaleup}~(c) shows the relative error of the ROM compared to the FOM solution,
\begin{equation}
    \epsilon = \left[\frac{\sum\limits_{m}^M \Vert \bq_m - \bPhi_{r(m)}\hat{\bq}_m \Vert^2_{\Omega_m}}{\sum\limits_{m}^M \Vert \bq_m\Vert^2_{\Omega_m}}\right]^{1/2}.
\end{equation}
Throughout all cases, the relative error of the ROM solution remains at about $1\%$.
While the accuracy of the ROM varies over the test cases,
its variance is not so large:
the largest variance was observed for $M=32^2$-component domain,
from $0.8\%$ to $3\%$.
This comparison clearly shows that
the component ROM can effectively make a fast and robust prediction at an extrapolated scale.
\par
We further demonstrate the robustness of the component ROM,
with a test problem that is qualitatively different from the one used for the sample generation.
We devised a right-hand side that has a sinusoidal wave along an Archimedean spiral
for a domain $\Omega = [0, L]^2$,
\begin{subequations}\label{eq:poisson-spiral}
    \begin{equation}
        f = \exp \left(-\frac{\vert r - \frac{s\theta L}{4\pi}\vert^2}{2w^2}\right)\cos 2\pi k\left\vert r - \frac{s\theta L}{4\pi}\right\vert,
    \end{equation}
    where $(r, \theta)$ are the polar coordinates of $\bx$ with respect to the center of $\Omega$,
    \begin{equation}
        \begin{split}
            r &= \vert\bx - (0.5L, 0.5L)\vert\\
            \theta &= \tan^{-1} \frac{x_2 - 0.5L}{x_1 - 0.5L}.
        \end{split}
    \end{equation}
    The entire boundary is set with the homogeneous Dirichlet condition,
    \begin{equation}
        g_{di} = 0,
    \end{equation}
    with $\partial\Omega \equiv \partial\Omega_{di}$.
\end{subequations}
Figure~\ref{fig:poisson-global}~(b) shows an example ROM solution of the spiral problem (\ref{eq:poisson-spiral})
on the global domain with $L=32$.
Such a spiral pattern was never seen from the samples,
so the solution must be extrapolated out of the basis constructed from the samples of the linear wave problem (\ref{eq:poisson-component}).
The same ROM system is tested for this spiral problem (\ref{eq:poisson-spiral})
over 100 test cases of parameter $(s, k)$ within a range of
\begin{equation}
s \in U[0, 0.7] \qquad k \in U[0, 0.7],
\end{equation}
with $L = 32$ and $w = 2$.
Even for this qualitatively different case,
the ROM was able to make robust predictions over all test cases.
In Figure~\ref{fig:poisson-scaleup}~(c),
the ROM error for the spiral problem case remains $\sim 0.7\%$ similar to the linear wave problem (\ref{eq:poisson-component}),
varying from $0.2\%$ to $6\%$.

\subsubsection{Performance with iterative solver}\label{subsubsec:poisson-iterative}
The use of a direct solver for a linear system is reasonable when 
the size of the system is not extremely large ($\dim(\bq) \gtrsim 10^7$)
or when an optimal preconditioner is not well-established yet.
In this work, we consider general PDE systems,
for which preconditioning strategies are not necessarily assumed to be well-developed,
so the direct solver is considered as the default linear system solver.
\par
However, as the problem size increases, the direct method's cost can be intractable in both memory and computation time.
This issue may also apply to ROM, although ROM tends to be less constrained by these limitations than FOM due to its reduced size.
In such cases, the selection of a preconditioner for ROM becomes crucial in achieving optimal performance
and should be given as much attention as the preconditioner for FOM.
While the direct solver is perfectly suitable for our demonstration,
we present the performance of an iterative solver for both FOM and ROM to illustrate how it scales with problem size.
This comparison envisions the potential necessity for an optimal preconditioner for ROM.
\par
\begin{figure}[htb]
    \begin{tikzpicture}[font=\small,]
    \begin{groupplot}[
        group style={
            group name = my plots,
            group size= 2 by 2,
            xlabels at =edge bottom,
            horizontal sep=2.5cm,
            vertical sep=2.2cm,
        },
        name=chung,
    ]    
\pgfplotsset{set layers=standard}%

        \nextgroupplot[
            height = 0.45\textwidth,
            width = 0.5\textwidth,
            xlabel={Number of components},
            ylabel={Assembly time ($s$)},
            tick scale binop ={\times},
            xmode=log, ymode=log,
            legend style={
                font=\small,
                draw=none, fill=none,
                at={(rel axis cs: 0., 1.0)},
                anchor=south west,
                nodes={scale=1.0},
                legend cell align={left},
                legend columns=4,
                /tikz/every even column/.append style={column sep=0.5cm},
            },
            legend image post style={mark options={scale=1.0, fill=white, line width=1.0}},
        ]

            \addlegendimage{
                line width=1.0,
                solid,
                mark=none,
                blue,
            }
            \addlegendentry{ROM (GS)}
            \addlegendimage{
                line width=1.0,
                solid,
                mark=none,
                red,
            }
            \addlegendentry{FOM (AMG)}
            \addlegendimage{
                line width=1.0,
                dashed,
                mark=none,
                blue,
            }
            \addlegendentry{ROM (Fig.~\ref{fig:poisson-scaleup})}
            \addlegendimage{
                line width=1.0,
                dashed,
                mark=none,
                red,
            }
            \addlegendentry{FOM (Fig.~\ref{fig:poisson-scaleup})}

            \addplot+ [
                line width=1.0,
                solid,
                mark=*,
                mark options={fill=white,},
                blue,
                error bars/.cd, y dir=both, y explicit,
            ]
            table [
                x expr=\thisrowno{0}^2, y index=1,
                y error minus expr=\thisrowno{1} - \thisrowno{4},
                y error plus expr=\thisrowno{5} - \thisrowno{1},
            ]{./poisson_prec_rom_assemble.txt};

            \addplot+ [
                line width=1.0,
                solid,
                mark=*,
                mark options={fill=white,},
                red,
                error bars/.cd, y dir=both, y explicit,
            ]
            table [
                x expr=\thisrowno{0}^2, y index=1,
                y error minus expr=\thisrowno{1} - \thisrowno{4},
                y error plus expr=\thisrowno{5} - \thisrowno{1},
            ]{./poisson_prec_fom_assemble.txt};

            \addplot+ [
                line width=1.0,
                dashed,
                mark=*,
                mark options={fill=white, solid},
                blue,
                error bars/.cd, y dir=both, y explicit,
            ]
            table [
                x expr=\thisrowno{0}^2, y index=1,
                y error minus expr=\thisrowno{1} - \thisrowno{4},
                y error plus expr=\thisrowno{5} - \thisrowno{1},
            ]{./poisson_direct_rom_assemble.txt};

            \addplot+ [
                line width=1.0,
                dashed,
                mark=*,
                mark options={fill=white, solid},
                red,
                error bars/.cd, y dir=both, y explicit,
            ]
            table [
                x expr=\thisrowno{0}^2, y index=1,
                y error minus expr=\thisrowno{1} - \thisrowno{4},
                y error plus expr=\thisrowno{5} - \thisrowno{1},
            ]{./poisson_direct_fom_assemble.txt};

        \nextgroupplot[
            height = 0.45\textwidth,
            width = 0.5\textwidth,
            xlabel={Number of components},
            ylabel={Solution time ($s$)},
            tick scale binop ={\times},
            xmode=log, ymode=log,
            legend style={
                font=\small,
                draw=none, fill=none,
                at={(rel axis cs: 0., 1.0)},
                anchor=south west,
                nodes={scale=1.0},
                legend cell align={left},
                legend columns=2,
                /tikz/every even column/.append style={column sep=0.5cm},
            },
            legend image post style={mark options={scale=1.0, fill=white, line width=1.0}},
        ]
        
            \addplot+ [
                line width=1.0,
                solid,
                mark=*,
                mark options={fill=white,},
                blue,
                select coords between index={0}{3},
                error bars/.cd, y dir=both, y explicit,
            ]
            table [
                x expr=\thisrowno{0}^2, y index=1,
                y error minus expr=\thisrowno{1} - \thisrowno{2},
                y error plus expr=\thisrowno{3} - \thisrowno{1},
            ]{./poisson_prec_rom_solve.txt};

            \addplot+ [
                line width=1.0,
                solid,
                mark=*,
                mark options={fill=white,},
                red,
                error bars/.cd, y dir=both, y explicit,
            ]
            table [
                x expr=\thisrowno{0}^2, y index=1,
                y error minus expr=\thisrowno{1} - \thisrowno{2},
                y error plus expr=\thisrowno{3} - \thisrowno{1},
            ]{./poisson_prec_fom_solve.txt};

            \addplot+ [
                line width=1.0,
                dashed,
                mark=*,
                mark options={fill=white, solid},
                blue,
                select coords between index={0}{3},
            ]
            table [x expr=\thisrowno{0}^2, y index=1,]{./poisson_direct_rom_solve.txt};

            \addplot+ [
                line width=1.0,
                dashed,
                mark=*,
                mark options={fill=white, solid},
                red,
                select coords between index={0}{3},
            ]
            table [x expr=\thisrowno{0}^2, y index=1,]{./poisson_direct_fom_solve.txt};

            \logLogSlopeTriangle{0.3}{0.1}{0.6}{1}{}{north}
            \logLogSlopeTriangle{0.6}{0.1}{0.45}{1.5}{}{north}

  \end{groupplot}
\node[below = 1.5cm of my plots c1r1.south west,
  anchor=west,
] {(a)};
\node[below = 1.5cm of my plots c2r1.south west,
  anchor=west,
] {(b)};
\end{tikzpicture}
%
    \caption{The comparison of (a) the assembly time and (b) the computation times between the ROM and the FOM for the Poisson component problem (\ref{eq:poisson-component}),
    using the conjugate gradient method with preconditioners (AMG for the FOM, and GS for the ROM).
    The marker denotes the median value of 100 test cases,
    and the error bar denotes $95\%$-confidence interval of the test cases.
    The results from Figure~\ref{fig:poisson-scaleup} are included as dashed lines for a reference to compare.}
    \label{fig:poisson-prec}
\end{figure}
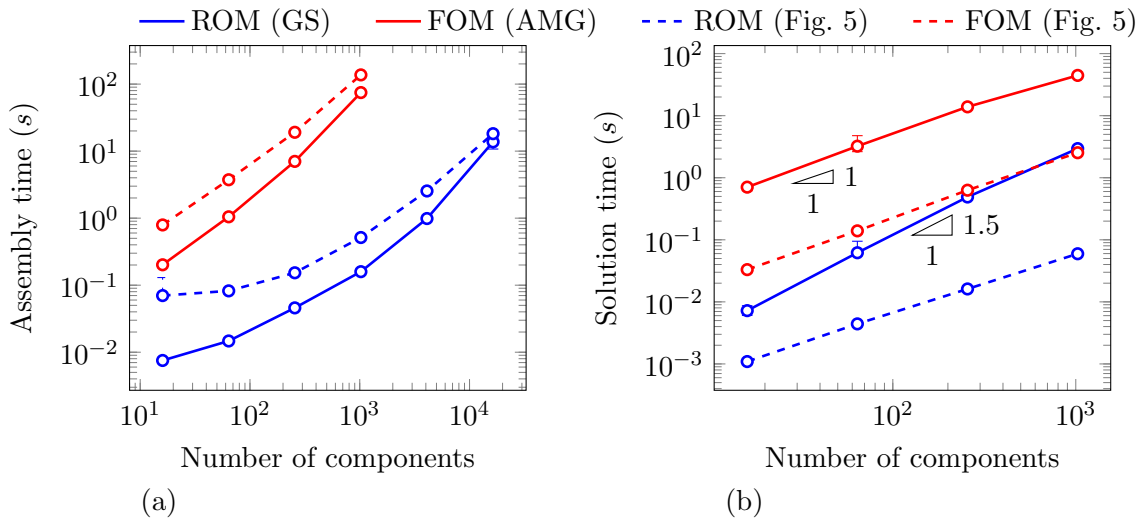
Figure~\ref{fig:poisson-prec} shows the computation time for the ROM and the FOM system using the iterative conjugate gradient method.
As the iterative solver does not require an LU factorization of the system \PR{like} the direct method,
the assembly time in Figure~\ref{fig:poisson-prec}~(a) was shorter than the direct method in Figure~\ref{fig:poisson-scaleup}~(b),
for both FOM and ROM.
However, this overhead cost for the direct method does not scale as fast as the matrix assembly itself,
so its portion in the assembly time decreases with the problem size.
\par
Figure~\ref{fig:poisson-prec}~(b) shows the computation time for the ROM and the FOM system, both using a preconditioner.
For the Poisson equation, various preconditioners have been developed,
among which multi-grid methods have shown their supremacy over all others, in particular algebraic multi-grid (AMG)~\cite{Trottenberg2000,Stuben2001}.
This is indeed the case for our FOM system as well.
For the FOM systems, though more expensive than the direct method,
AMG efficiently solves the system with the complexity of $\cO(M)$.
The same AMG, however, does not necessarily accelerate the ROM system.
Using other conventional preconditioners, such as symmetric Gauss-Seidel (GS) method,
slightly improves the computation time, solving 100 times faster than the FOM with the AMG preconditioner.
However, the scaling behavior remains $\cO(M^{1.5})$, and the computation time gap between FOM and ROM decreases with the problem size.
Although the ROM still solves the problem about 20 times faster than the FOM at our largest test case,
its advantage over FOM decreases with the problem size.
\par
The results in Figure~\ref{fig:poisson-prec} show a need for developing preconditioners for the ROM system.
One main reason for the ineffectiveness of the conventional preconditioners on the ROM system
is the sparse-block matrix structure of the ROM system:   
the system (\ref{eq:weak-gov-reduced}) is constructed with sparse non-zero block matrices,
where each block matrix is dense based on the ROM operators (\ref{eq:comp-bilinear-mat-reduced}).
This is qualitatively different from what the conventional preconditioners are designed for.
\par
A caveat in developing a preconditioner is that the effectiveness of a preconditioner is often problem-dependent.
\PR{In many cases}, an optimal preconditioner for a given PDE system is not simple to implement or well-established throughout the community.
It also often involves prior physics knowledge for the given system to develop an optimal preconditioner~\cite{Kay2002,Elman2014,Farrell2019},
which is not always available.
In such cases, the direct method is still a viable choice over a wide range of the problem size.
This is even more so for ROM with its greatly reduced system size.
Our goal in this research is to present a reduced order modeling approach for general PDE problems.
Therefore, we did not further explore the options of developing an effective preconditioner for each problem,
which is beyond the scope of our study.

\subsubsection{Performance scaling with the number of basis}\label{subsubsec:poisson-scaling}
We further investigate the performance of the ROM depending on basis dimension $R_1$.
For demonstration, the Poisson component problem (\ref{eq:poisson-component}) is solved for $\Omega = [0, 4]^2$
with $M=4^2$ components, at a fixed parameter value
\begin{equation}
    \bk = (0.6, -0.6)
    \qquad
    \bk_b = (0.05, 0.04)
    \qquad
    \theta = 0.1
    \qquad
    \theta_b = 0.
\end{equation}
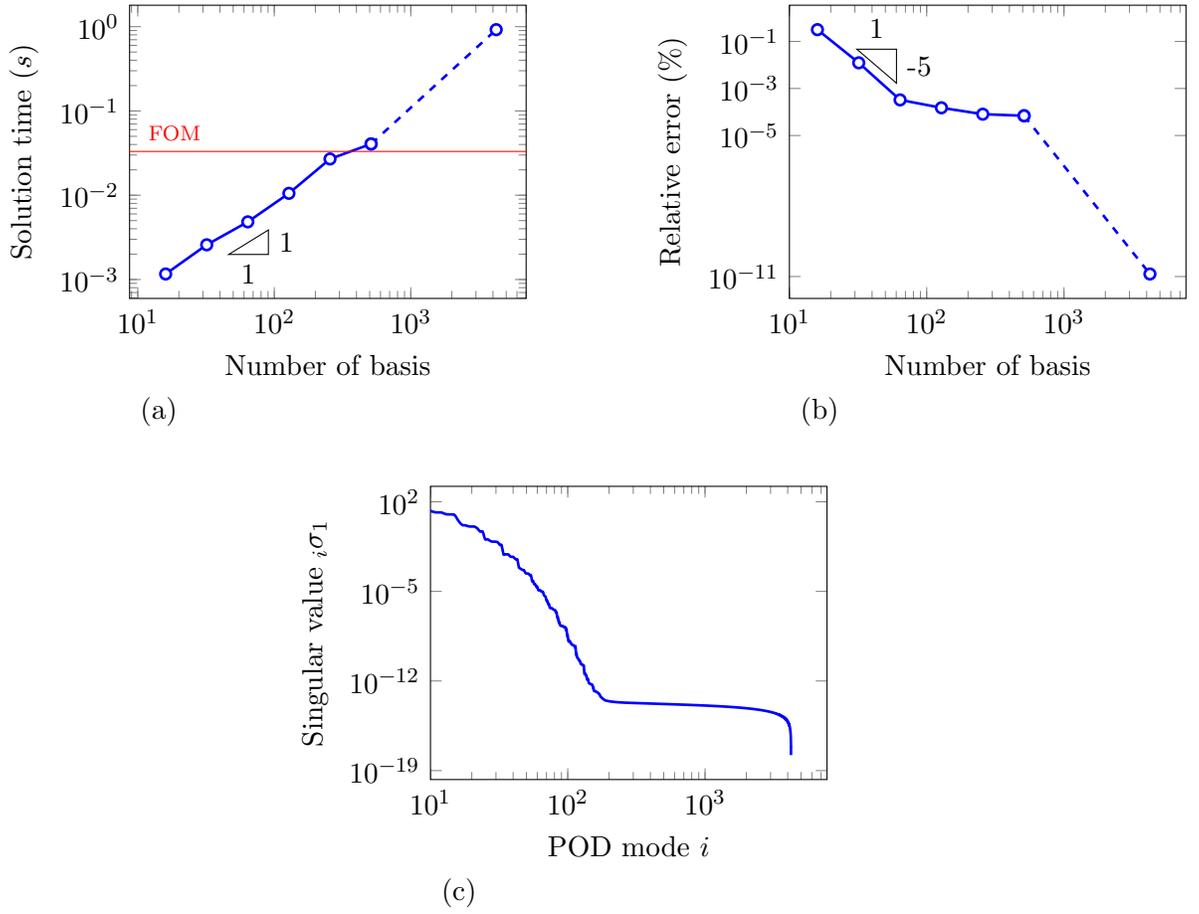
\begin{figure}[tbhp]
    \begin{tikzpicture}[font=\small,]
    \begin{groupplot}[
        group style={
            group name = my plots,
            group size= 2 by 2,
            xlabels at =edge bottom,
            horizontal sep=3.5cm,
            vertical sep=2.5cm,
        },
        name=chung,
    ]    
\pgfplotsset{set layers=standard}%

        \nextgroupplot[
            height = 0.4\textwidth,
            width = 0.5\textwidth,
            xlabel={Number of basis},
            ylabel={Solution time ($s$)},
            tick scale binop ={\times},
            xmode=log, ymode=log,
            xmax=7e3,
            legend style={
                font=\small,
                draw=none, fill=none,
                at={(rel axis cs: 0., 1.0)},
                anchor=south west,
                nodes={scale=1.0},
                legend cell align={left},
                legend columns=4,
                /tikz/every even column/.append style={column sep=0.5cm},
            },
            legend image post style={mark options={scale=1.0, fill=white, line width=1.0}},
        ]
        


            \addplot+ [
                line width=1.0,
                solid,
                mark=*,
                mark options={fill=white,},
                blue,
                select coords between index={0}{5},
            ]
            table [x index=0, y index=2]{./poisson_component.direct.txt};

            \addplot+ [
                line width=1.0,
                dashed,
                mark=*,
                mark options={fill=white, solid},
                blue,
                select coords between index={5}{6},
            ]
            table [x index=0, y index=2]{./poisson_component.direct.txt};

            \draw[red, solid,] (axis cs: 1e0, 3.30898E-02) -- (axis cs: 1e4, 3.30898E-02);
            \node[anchor=south west, font=\scriptsize, red] at (axis cs: 1e1, 3.30898E-02) {FOM};

            \logLogSlopeTriangle{0.35}{0.1}{0.15}{1}{}{north}

        \nextgroupplot[
            height = 0.4\textwidth,
            width = 0.5\textwidth,
            xlabel={Number of basis},
            ylabel={Relative error ($\%$)},
            tick scale binop ={\times},
            xmode=log, ymode=log,
            xmin=1e1,
            ytick={1e-11, 1e-5, 1e-3, 1e-1, 1e1},
            legend style={
                font=\small,
                draw=none, fill=none,
                at={(rel axis cs: 0., 1.0)},
                anchor=south west,
                nodes={scale=1.0},
                legend cell align={left},
                legend columns=4,
                /tikz/every even column/.append style={column sep=0.5cm},
            },
            legend image post style={mark options={scale=1.0, fill=white, line width=1.0}},
        ]
        
            \addplot+ [
                line width=1.0,
                solid,
                mark=*,
                mark options={fill=white,},
                blue,
                select coords between index={0}{5},
            ]
            table [x index=0, y expr=\thisrowno{5} * 1e2]{./poisson_component.direct.txt};

            \addplot+ [
                line width=1.0,
                dashed,
                mark=*,
                mark options={fill=white, solid},
                blue,
                select coords between index={5}{6},
            ]
            table [x index=0, y expr=\thisrowno{5} * 1e2]{./poisson_component.direct.txt};

    
            \logLogSlopeTriangle{0.27}{0.1}{0.85}{-5}{}{south}

        \nextgroupplot[
            height = 0.4\textwidth,
            width = 0.5\textwidth,
            ylabel={Singular value ${}_i\sigma_1$},
            xlabel={POD mode $i$},
            tick scale binop ={\times},
            xmode=log, ymode=log,
            xmin=1e1,
            legend style={
                font=\small,
                draw=none, fill=none,
                at={(rel axis cs: 0., 1.0)},
                anchor=south west,
                nodes={scale=1.0},
                legend cell align={left},
                legend columns=4,
                /tikz/every even column/.append style={column sep=0.5cm},
            },
            legend image post style={mark options={scale=1.0, fill=white, line width=1.0}},
            xshift=4cm,
        ]
        
            \addplot+ [
                line width=1.0,
                smooth, solid,
                mark=none,
            ]
            table [x expr=\coordindex+1, y index=0]{./poisson_component_square_sv.txt};
 
\end{groupplot}
\node[below = 1.5cm of my plots c1r1.south west,
    anchor=west,
] {(a)};
\node[below = 1.5cm of my plots c2r1.south west,
    anchor=west,
] {(b)};
\node[below = 1.5cm of my plots c1r2.south west,
    anchor=west, xshift=4cm,
] {(c)};
\end{tikzpicture}
%
    \caption{Performance scaling of the ROM with respect to the  basis dimension $R_1$:
    (a) the computation time of the resulting ROM system;
    (b) the relative error of the ROM solution; and
    (c) the full spectrum of the singular values of the POD basis $\Omega_1$ in Figure~\ref{fig:poisson-sample}(b).
    The computation time for the FOM system is included as a reference to compare.
    }
    \label{fig:poisson-scaling}
\end{figure}
Figure~\ref{fig:poisson-scaling}(a) shows the computation time of the resulting ROM system,
depending on basis dimension.
The ROM system is solved via the same direct LU method.
The computation time scales linearly as $\cO(R_1)$.
Due to this scaling, the ROM becomes as slow as the FOM only with a 200 basis dimension,
however, its accuracy also scales rapidly with $R_1$.
In Figure~\ref{fig:poisson-scaling}(b),
the ROM already achieves below $1\%$ error only with 16 basis vectors.
Furthermore, the relative error of the ROM solution decays as $\cO(R_1^{-5})$,
much faster than the computation time.
Considering that the ROM typically targets for a moderate accuracy of $0.1\%\sim 1\%$,
this result implies that
the proposed component ROM can achieve the desired accuracy only with the first few basis vectors.
\par
While $R_1 > 100$ would be beyond the practical range for ROM,
we further investigated the scaling behavior to provide an insight about the linear-subspace ROM and sample generation.
In Figure~\ref{fig:poisson-scaling}(b),
the relative error seems to stagnate around $10^{-4}\%$ for $R_1 > 10^2$.
While there seems to be a constant factor of error,
this is in fact related to how the linear subspace $\bPhi$ is identified.
Figure~\ref{fig:poisson-scaling}(c)
shows the full spectrum of the singular values of the POD basis $\tilde{\bPhi}_1$.
The singular values decrease rapidly down to a negligible value of $10^{-12}$ around the 200th basis vector,
indicating that the sample snapshot $\bQ_1$ can be perfectly spanned only with the first 200 POD basis vectors.
The rest of the basis (about 4000 basis vectors) is essentially random orthonormal vectors corresponding to the null space of the snapshot $\bQ_1$,
which are not necessarily well factorized to represent the solution.
When the prediction is outside the training data,
a non-negligible portion of the solution may not be spanned by the first 200 basis vectors,
where then the remaining 4000 basis vectors are used to represent this portion.
However, since these 4000 basis vectors are not well factorized,
this non-negligible portion requires essentially the entire basis in order to be effectively represented.
This is confirmed in Figure~\ref{fig:poisson-scaling}~(b) by checking the accuracy of the ROM with $R_1 = 4225 \equiv N_1$.
The relative error, which was seemingly stagnated previously, decreases down to a negligible value of $10^{-12}$.
\par
This behavior shows an implicit impact of the samples on the resulting ROM performance.
The richer the samples we collect for identifying the linear subspace,
the broader the spectrum of the effective basis becomes,
and therefore, the smaller error the ROM can achieve as it increases the basis dimension.

\section{Stokes flow equation}\label{sec:stokes}

\subsection{Governing equation}
The governing equation (\ref{eq:gov}) for Stokes flow is defined for
$\tbq\equiv (\tilde{\bu}, \tilde{p})$ with $\tilde{\bu}\in H^1(\Omega)^{\dim(\Omega)}$, $\tilde{p} \in H^1(\Omega)$ and $\f\equiv 0$,
\begin{subequations}\label{eq:stokes-gov}
    \begin{equation}
        -\nu\nabla^2 \tilde{\bu} + \nabla\tilde{p} = 0 \qquad \text{in } \Omega,
    \end{equation}
    \begin{equation}
        \nabla\cdot\tilde{\bu} = 0 \qquad \text{in } \Omega,
    \end{equation}
    with boundary conditions on $\partial\Omega$,
    \begin{equation}
        \tilde{\bu} = \bg_{di} \qquad \text{on } \partial\Omega_{di},
    \end{equation}
    \begin{equation}
        \bn\cdot(-\nu\nabla\tilde{\bu} + \tilde{p}\bI) = \bg_{ne} \qquad \text{on } \partial\Omega_{ne},
    \end{equation}
\end{subequations}
where $\bu$ and $p$ denotes the velocity and pressure field respectively,
and $\nu\equiv\frac{\mu}{\mu_0}=1.1$ is the non-dimensionalized dynamic viscosity with respect to a reference value.
\todo{
    KC: realized non-dimensionalized Stokes flow wouldn't have this coefficient, unless we consider a reference value.
    We can either do this or just keep the dimensional units?
}
(\ref{eq:stokes-gov}) describes the conservation of momentum with incompressibility condition.
(\ref{eq:stokes-gov}) is decomposed into subdomains as in (\ref{eq:dd-gov}),
\begin{subequations}\label{eq:stokes-dd-gov}
    \begin{equation}
        -\nu\nabla^2 \tilde{\bu}_m + \nabla\tilde{p}_m = 0 \qquad \text{in } \Omega_m,
    \end{equation}
    \begin{equation}
        \nabla\cdot\tilde{\bu}_m = 0 \qquad \text{in } \Omega_m,
    \end{equation}
    with interface constraints on $\Gamma_{m,n}$,
    \begin{equation}
        \jump{\tilde{\bu}} = \avg{\bn\cdot\nabla\tilde{\bu}} = 0,
    \end{equation}
    \begin{equation}
        \jump{\tilde{p}} = \avg{\bn\cdot\nabla\tilde{p}} = 0.
    \end{equation}
\end{subequations}

\subsection{Weak form finite element formulation}\label{subsec:stokes-dg}
Several choices of DG finite element spaces for Stokes flow are available
for our domain decomposition framework~\cite{Toselli2002,Cockburn2002,Elman2014}.
Our choice of the finite element space (\ref{eq:fes}) is Taylor-Hood elements~\cite{Taylor1973},
\begin{subequations}
    \begin{equation}\label{eq:stokes-Qms}
        \bq_m \equiv (\bu_m, p_m) \in \rQ_{m, s} = \rU_{m, s+1} \otimes \rP_{m,s},
    \end{equation}
    where the finite element spaces for velocity and pressure are defined respectively,
    \begin{equation}
        \rU_{m, s} = \left\{ \bu_m \in H^1(\Omega_m) \;\bigg|\; \bu_m\big|_{\kappa} \in V_s(\kappa)^{\dim(\Omega)} \quad \forall \kappa\in\cT(\Omega_m) \right\}
    \end{equation}
    \begin{equation}
        \rP_{m, s} = \left\{ p_m \in H^1(\Omega_m) \cap H^1_0(\Omega) \;\bigg|\; p_m\big|_{\kappa} \in V_s(\kappa) \quad \forall \kappa\in\cT(\Omega_m) \right\},
    \end{equation}
\end{subequations}
where $\bq_m$ is discretized in a continuous Galerkin way as for the Poisson equation in Section~\ref{sec:poisson}.
In (\ref{eq:stokes-Qms}), the velocity space is defined one-order higher than the pressure space
in order to satisfy the Ladyzhenskaya-Babu\v{s}ka-Brezzi (LBB) condition for existence of a solution~\cite{Babuvska1971,Brezzi1974,Ladyzhenskaya1963,Taylor1973}.
In the case of $\partial\Omega_{ne}=\varnothing$, the pressure is restricted to have mean value zero in the global domain $\Omega$.
\par
The weak form of (\ref{eq:stokes-dd-gov}) corresponding to (\ref{eq:weak-gov})
is formulated with respect to $\bqs=\{(\bus_m, \ps_m)\}$ and $\bq=\{(\bu_m, p_m)\}$.
The bilinear-form operators in (\ref{eq:dd-gov-op}) are defined as a block matrix form,
\begin{subequations}
    \begin{equation}
        \form{B_m}{\bqs}{\cL[\bq]} =
        \begin{pmatrix}
            \form{a}{\bus}{\bu}_{m} & \form{b}{\bus}{p}_{m}\\
            \form{b}{\bu}{\ps}_{m} & 0 \\
        \end{pmatrix},
    \end{equation}
    where each block corresponds to
    \begin{equation}
        \form{a}{\bus}{\bu}_{m} = \inprod{\nu\nabla\bus_m}{\nabla\bu_m}_{\Omega_m}
    \end{equation}
    \begin{equation}
        \form{b}{\bus}{p}_{m} = -\inprod{\nabla\cdot\bus_m}{p_m}_{\Omega_m}.
    \end{equation}
    Likewise, the interface bilinear weak form operator is
    \begin{equation}
        \form{B_{m,n}}{\bqs}{\cL[\bq]} =
        \begin{pmatrix}
            \form{a}{\bus}{\bu}_{m,n} & \form{b}{\bus}{p}_{m,n}\\
            \form{b}{\bu}{\ps}_{m,n} & 0 \\
        \end{pmatrix},
    \end{equation}
    with its blocks
    \begin{equation}
    \begin{split}
        \form{a}{\bus}{\bu}_{m,n} =&
        - \inprod{\avg{\bn\cdot\nu\nabla\bus}}{\jump{\bu}}_{\Gamma_{m,n}}\\
        &- \inprod{\jump{\bus}}{\avg{\bn\cdot\nu\nabla\bu}}_{\Gamma_{m,n}}\\
        &+ \inprod{\gamma\Delta\bx^{-1}\jump{\bus}}{\jump{\bu}}_{\Gamma_{m,n}}
    \end{split}
    \end{equation}
    \begin{equation}
        \form{b}{\bus}{p}_{m,n} = \inprod{\jump{\bn\cdot\bus}}{\avg{p}}_{\Gamma_{m,n}}.
    \end{equation}
    The boundary bilinear form on $\Gamma_{m,di}$ the Dirichlet boundary of the subdomain is
    \begin{equation}
        \form{B_{m,di}}{\bqs}{\cL[\bq]} =
        \begin{pmatrix}
            \form{a}{\bus}{\bu}_{m,di} & \form{b}{\bus}{p}_{m,di}\\
            \form{b}{\bu}{\ps}_{m,di} & 0 \\
        \end{pmatrix},
    \end{equation}
    with its blocks
    \begin{equation}
    \begin{split}
        \form{a}{\bus}{\bu}_{m,di} =&
        - \inprod{\bn\cdot\nu\nabla\bus}{\bu}_{\Gamma_{m,di}}\\
        &- \inprod{\bus}{\bn\cdot\nu\nabla\bu}_{\Gamma_{m,di}}\\
        &+ \inprod{\gamma\Delta\bx^{-1}\bus}{\bu}_{\Gamma_{m,di}}
    \end{split}
    \end{equation}
    \begin{equation}
        \form{b}{\bus}{p}_{m,di} = \inprod{\bn\cdot\bus}{p}_{\Gamma_{m,di}}.
    \end{equation}
\end{subequations}
With $\form{L_m}{\bqs}{\f} = 0$, we only have the boundary linear weak form operators,
\begin{subequations}
    \begin{equation}
        \form{L_{ne}}{\bqs}{\f} =
        \begin{pmatrix}
            \inprod{\bus}{\bg_{ne}}_{\partial\Omega_{ne}}\\
            0
        \end{pmatrix}
    \end{equation}
    \begin{equation}
        \form{L_{di}}{\bqs}{\f} =
        \begin{pmatrix}
            \inprod{\bus}{\gamma\Delta\bx^{-1}\bg_{di}}_{\partial\Omega_{di}}
            + \inprod{\bn\cdot\nu\nabla\bus}{\bg_{di}}_{\partial\Omega_{di}}\\
            \inprod{\ps}{\bn\cdot\bg_{di}}_{\partial\Omega_{di}}.
        \end{pmatrix}
    \end{equation}
\end{subequations}
The interface penalty strength $\gamma = \nu(s+1)^2$ is used for $\rQ_{m,s}$
to ensure the stability of the solution~\cite{Toselli2002}.
\par

\subsection{Component-level samples for linear subspace training}
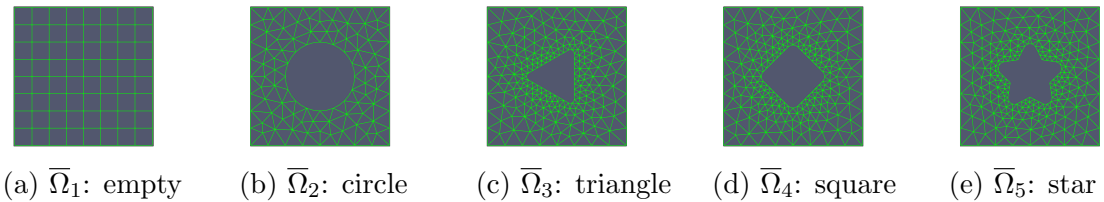
\begin{figure}
    \begin{tikzpicture}[font=\small,]
    \begin{groupplot}[
        group style={
            group name = my plots,
            group size= 5 by 1,
            xlabels at =edge bottom,
            horizontal sep=1.3cm,
            vertical sep=2.2cm,
        },
        name=chung,
        height = 0.25\textwidth,
        width = 0.25\textwidth,
    ]    
\pgfplotsset{set layers=standard}%

        \nextgroupplot[
            xtick=\empty, ytick=\empty,
            tick scale binop ={\times},
            xmin = 0, xmax = 1,
            ymin = 0, ymax = 1,
        ]
        
            \edef\imagepath{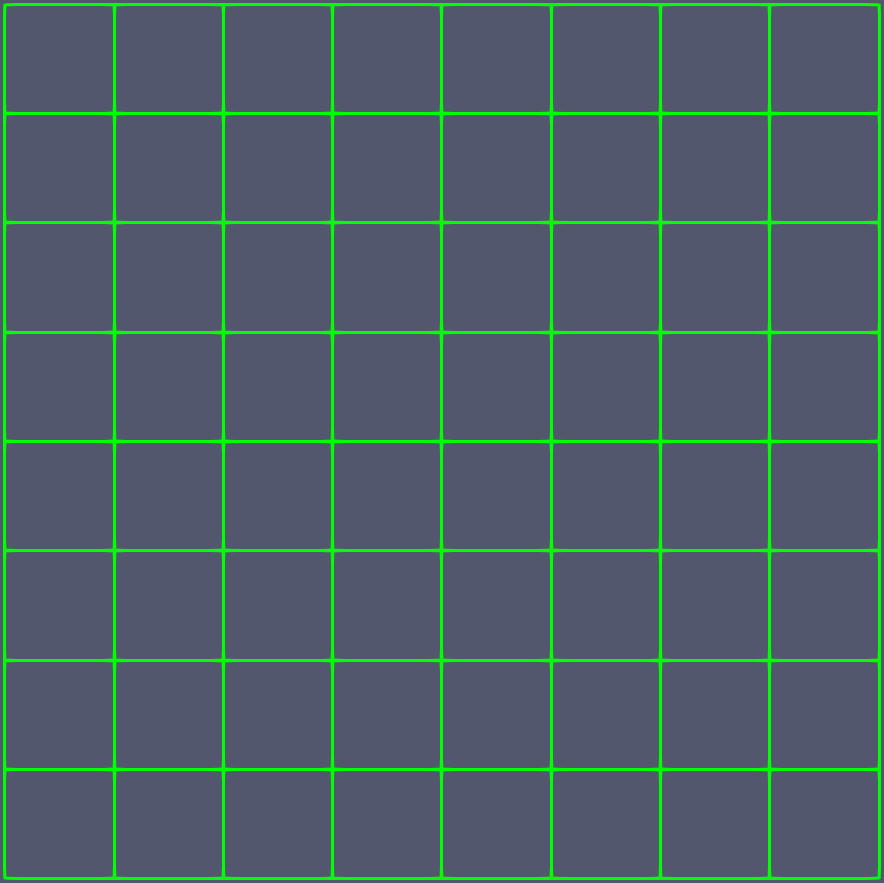}
            \addplot graphics[xmin=0,xmax=1,ymin=0,ymax=1]{\imagepath};

        \nextgroupplot[
            xtick=\empty, ytick=\empty,
            tick scale binop ={\times},
            xmin = 0, xmax = 1,
            ymin = 0, ymax = 1,
        ]
        
            \edef\imagepath{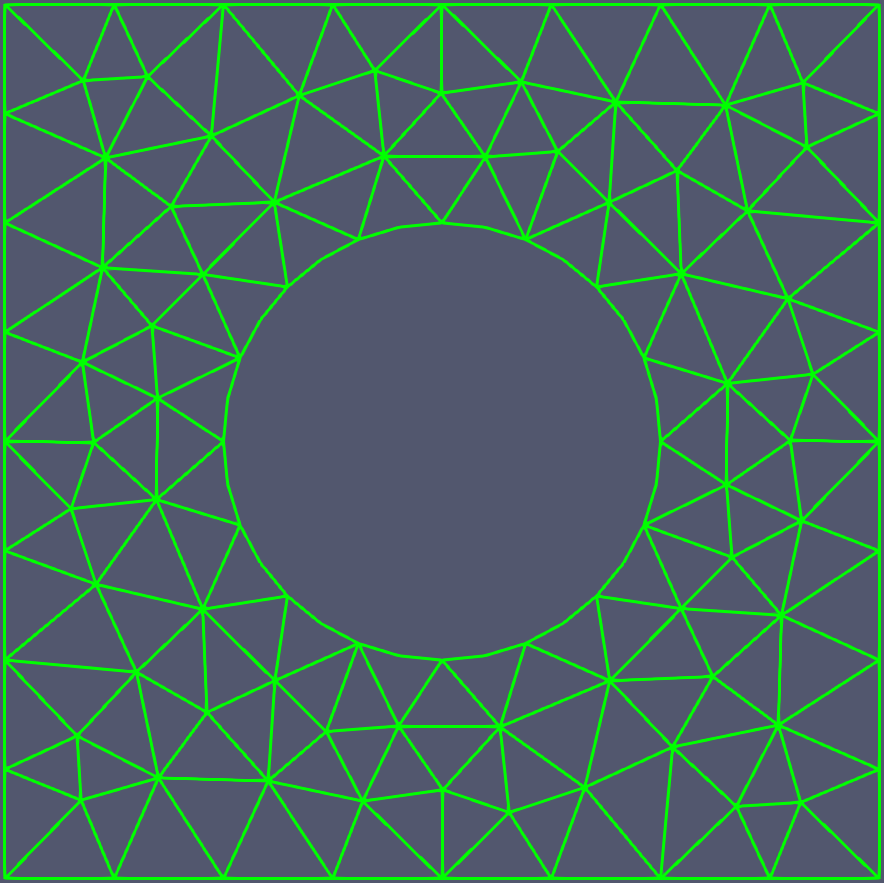}
            \addplot graphics[xmin=0,xmax=1,ymin=0,ymax=1]{\imagepath};
        
        \nextgroupplot[
            xtick=\empty, ytick=\empty,
            tick scale binop ={\times},
            xmin = 0, xmax = 1,
            ymin = 0, ymax = 1,
        ]
        
            \edef\imagepath{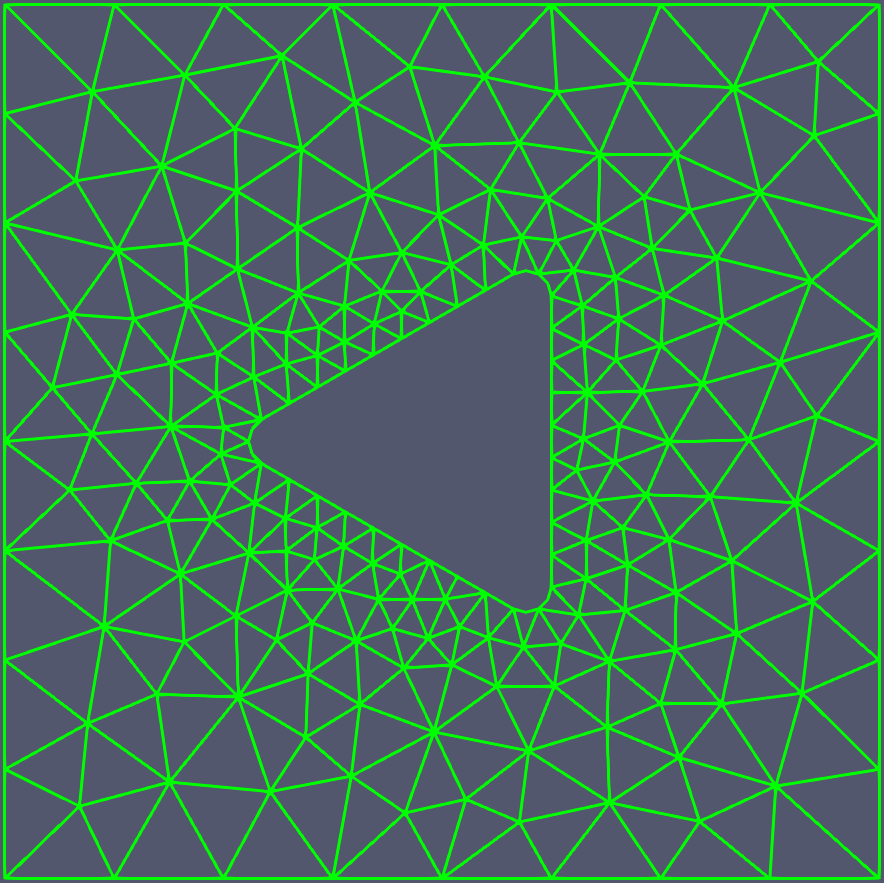}
            \addplot graphics[xmin=0,xmax=1,ymin=0,ymax=1]{\imagepath};

        \nextgroupplot[
            xtick=\empty, ytick=\empty,
            tick scale binop ={\times},
            xmin = 0, xmax = 1,
            ymin = 0, ymax = 1,
        ]
        
            \edef\imagepath{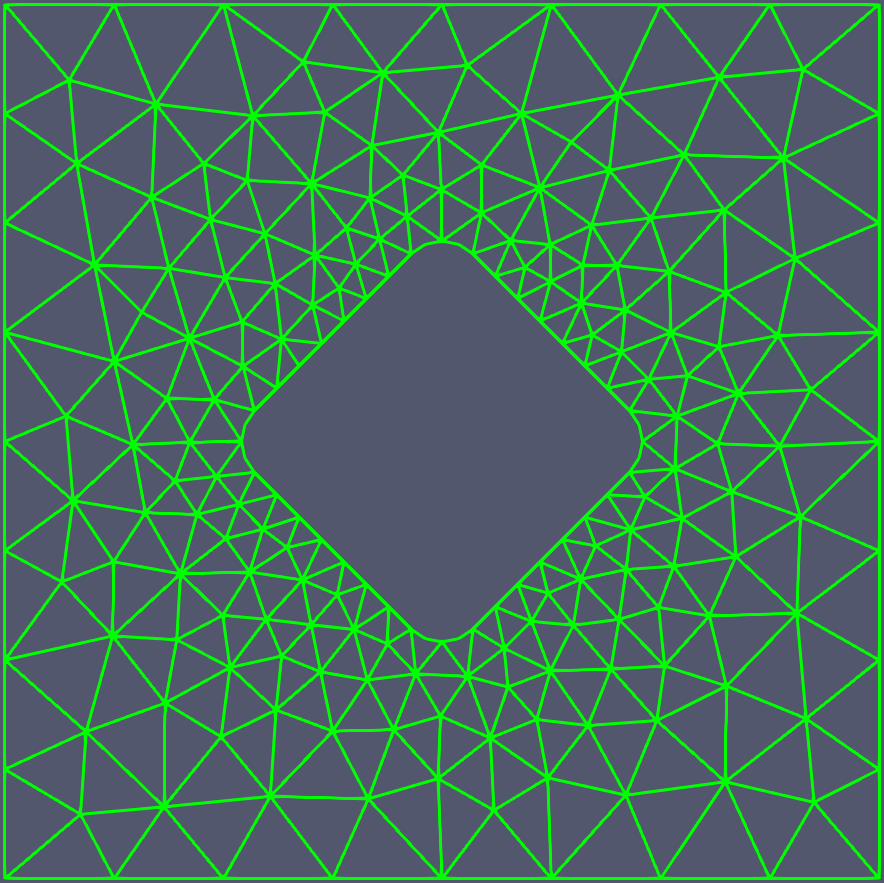}
            \addplot graphics[xmin=0,xmax=1,ymin=0,ymax=1]{\imagepath};

        \nextgroupplot[
            xtick=\empty, ytick=\empty,
            tick scale binop ={\times},
            xmin = 0, xmax = 1,
            ymin = 0, ymax = 1,
        ]
        
            \edef\imagepath{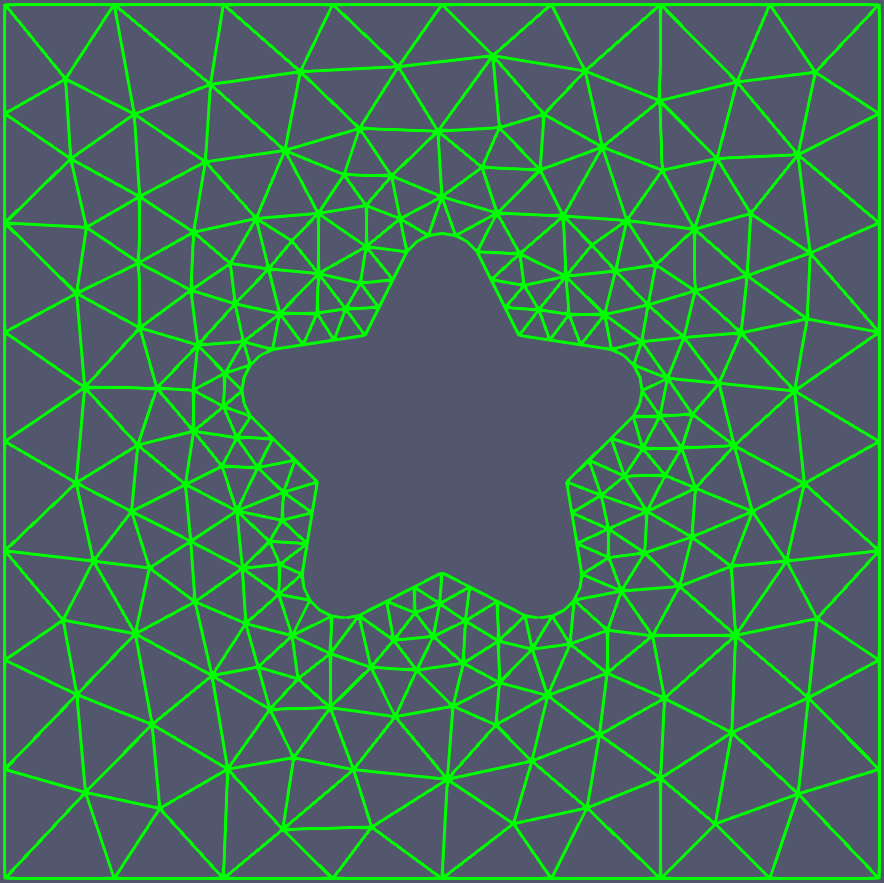}
            \addplot graphics[xmin=0,xmax=1,ymin=0,ymax=1]{\imagepath};

  \end{groupplot}
\node[below = .5cm of my plots c1r1.south west,
    anchor=west, xshift=-.3cm,
] {(a) $\bOmega_1$: empty};
\node[below = .5cm of my plots c2r1.south west,
    anchor=west, xshift=-.3cm,
] {(b) $\bOmega_2$: circle};
\node[below = .5cm of my plots c3r1.south west,
    anchor=west, xshift=-.3cm,
] {(c) $\bOmega_3$: triangle};
\node[below = .5cm of my plots c4r1.south west,
    anchor=west, xshift=-.3cm,
] {(d) $\bOmega_4$: square};
\node[below = .5cm of my plots c5r1.south west,
    anchor=west, xshift=-.3cm,
] {(e) $\bOmega_5$: star};
\end{tikzpicture}
%
    \caption{Reference domains used as components:
        (a) empty,
        (b) circle,
        (c) triangle,
        (d) square,
        and (e) star.
    }
    \label{fig:stokes-comp}
\end{figure}

For this demonstration, five different reference domains $\rC =\{\bOmega_1, \ldots, \bOmega_5\}$ are
considered as components for building up the global-scale system.
All reference domains lie within a unit square $\bOmega_r \subset [0, 1]^2$,
with an obstacle (or none) within them, as shown in Figure~\ref{fig:stokes-comp}.
\todo[inline]{TYL: Do we know how results might change with mesh size? For example, for the flow past the star, there are probably flow features occuring in the interior corners of the star that are only observed if the mesh is sufficiently fine. Is there any reason to think that the number of basis vectors for some target accuracy increases with decreasing mesh size?}
The type and number of elements of each component is summarized in Table~\ref{tab:stokes-comp}.
\begin{table}[thbp]
    \centering
    \begin{tabular}{| c | c | c | c | c |}
        \hline
        $\bOmega_r$ (name) & Element type & Number of elements & $N_r$ & $S_r$ \\\hline
        $\bOmega_1$ (empty) & quadrilateral & 64 & 1539 & 1115 \\\hline
        $\bOmega_2$ (circle) & triangular & 168 & 2040 & 1076 \\\hline
        $\bOmega_3$ (triangle) & triangular & 393 & 4583 & 1098 \\\hline
        $\bOmega_4$ (square) & triangular & 372 & 4364 & 1114 \\\hline
        $\bOmega_5$ (star) & triangular & 415 & 4873 & 1197 \\\hline
    \end{tabular}
    \caption{Specification of component reference domains.}
    \label{tab:stokes-comp}
\end{table}
\par
\begin{figure}
    \begin{tikzpicture}[font=\small,]
    \begin{groupplot}[
        group style={
            group name = my plots,
            group size= 2 by 2,
            xlabels at =edge bottom,
            horizontal sep=3.5cm,
            vertical sep=2.2cm,
        },
        name=chung,
    ]    
\pgfplotsset{set layers=standard}%

        \nextgroupplot[
            height = 0.45\textwidth,
            width = 0.45\textwidth,
            ylabel={$x_2$},
            xlabel={$x_1$},
            tick scale binop ={\times},
            xmin = 0, xmax = 2,
            ymin = 0, ymax = 2,
            point meta min=0.0, point meta max=1.4,
            colorbar, colormap/viridis,
            colorbar style={
                font=\scriptsize,
                xticklabel pos=upper,
                scaled y ticks=false,
                /pgf/number format/precision=4,
                at={(rel axis cs: 1.01, 0.)}, anchor=south west,
                xlabel=$\vert\bu\vert(\bx)$,
            }
        ]
        
            \edef\imagepath{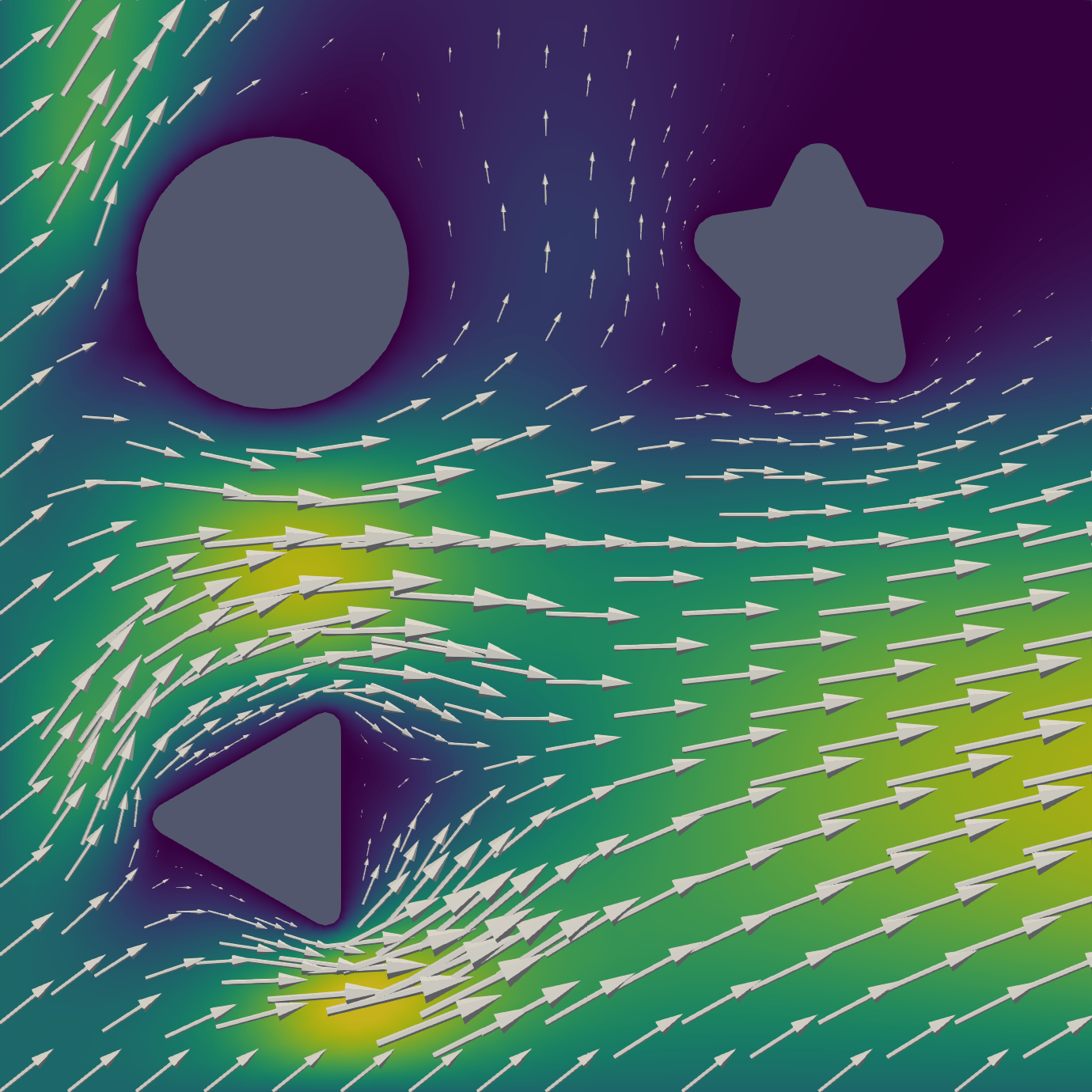}
            \addplot graphics[xmin=0.,xmax=2.,ymin=0.,ymax=2.]{\imagepath};

            \draw[
                red,
                dashed,
            ] (axis cs: 0, 1) -- (axis cs: 2, 1);
            \draw[
                red,
                dashed,
            ] (axis cs: 1, 0) -- (axis cs: 1, 2);

        \nextgroupplot[
            height = 0.45\textwidth,
            width = 0.45\textwidth,
            ylabel={$x_2$},
            xlabel={$x_1$},
            tick scale binop ={\times},
            xmin = 0, xmax = 2,
            ymin = 0, ymax = 2,
            point meta min=0.0, point meta max=1.5,
            colorbar, colormap/viridis,
            colorbar style={
                font=\scriptsize,
                xticklabel pos=upper,
                scaled y ticks=false,
                /pgf/number format/precision=4,
                at={(rel axis cs: 2.8, 0.)}, anchor=south west,
                xlabel=$\vert\bu\vert(\bx)$,
            }
        ]
        
            \edef\imagepath{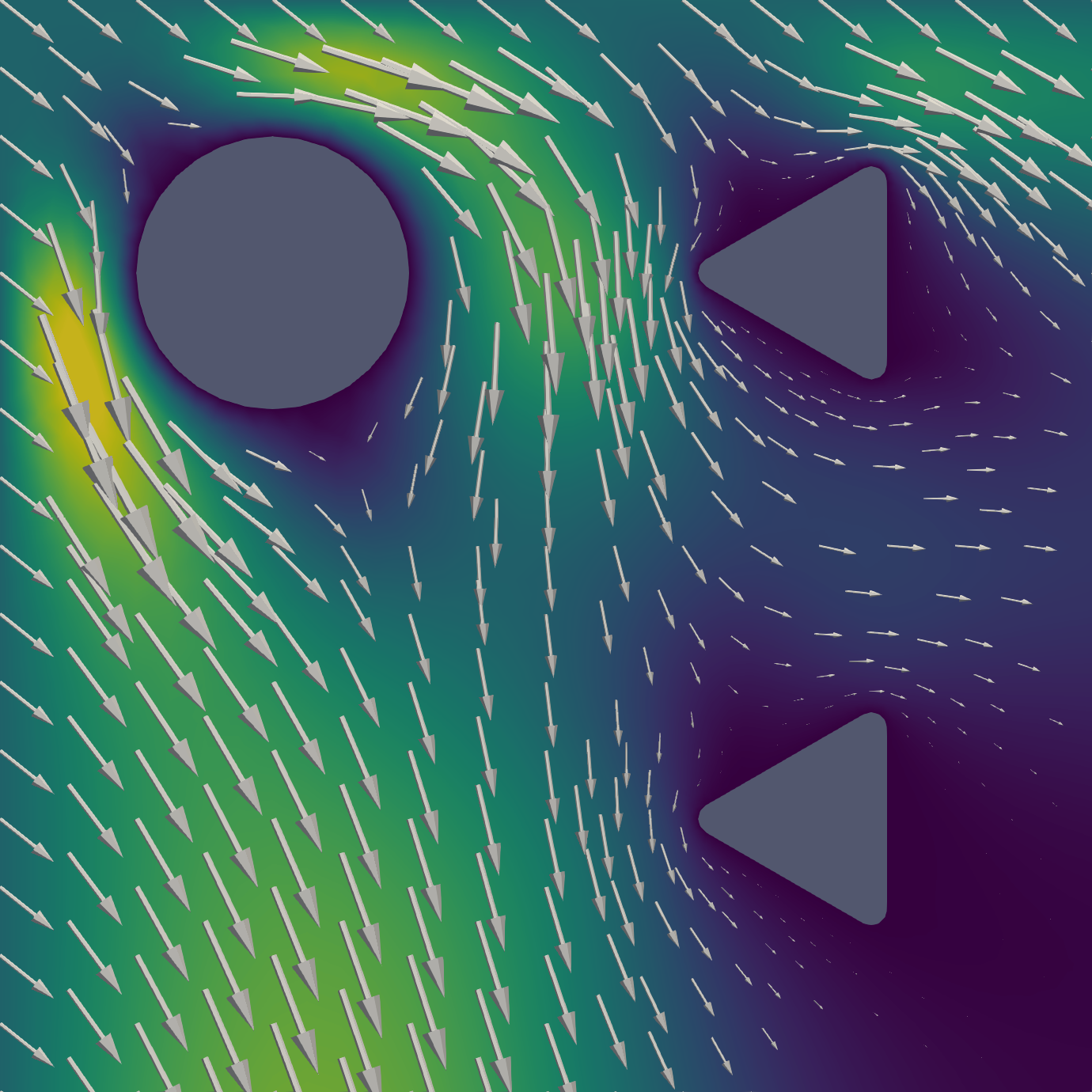}
            \addplot graphics[xmin=0.,xmax=2.,ymin=0.,ymax=2.]{\imagepath};

            \draw[
                red,
                dashed,
            ] (axis cs: 0, 1) -- (axis cs: 2, 1);
            \draw[
                red,
                dashed,
            ] (axis cs: 1, 0) -- (axis cs: 1, 2);

        \nextgroupplot[
            height = 0.45\textwidth,
            width = 0.5\textwidth,
            ylabel={Singular value ${}_i\sigma_r$},
            xlabel={POD mode $i$},
            tick scale binop ={\times},
            ymode=log,
            xmin=0, xmax=1000,
            legend style={
                draw=none, fill=none,
                legend cell align={left},
                at={(rel axis cs: 1., 1.0)},
                anchor=north west,
            },
            xshift=4cm,
        ]
        
            \addplot+ [
                line width=0.5,
                smooth, solid,
                mark=none,
            ]
            table [x expr=\coordindex+1, y index=0]{./stokes_basis_empty_sv.txt};
            \addplot+ [
                line width=0.5,
                smooth, solid,
                mark=none,
            ]
            table [x expr=\coordindex+1, y index=0]{./stokes_basis_square-circle_sv.txt};
            \addplot+ [
                line width=0.5,
                smooth, solid,
                mark=none,
            ]
            table [x expr=\coordindex+1, y index=0]{./stokes_basis_square-triangle_sv.txt};
            \addplot+ [
                line width=0.5,
                smooth, solid,
                mark=none,
            ]
            table [x expr=\coordindex+1, y index=0]{./stokes_basis_square-square_sv.txt};
            \addplot+ [
                line width=0.5,
                brown,
                smooth, solid,
                mark=none,
            ]
            table [x expr=\coordindex+1, y index=0]{./stokes_basis_square-star_sv.txt};

            \legend{$r=1$ empty, $r=2$ circle, $r=3$ triangle, $r=4$ square, $r=5$ star}
 
  \end{groupplot}
\node[below = 1.5cm of my plots c1r1.south west,
    anchor=west,
] {(a)};
\node[below = 1.5cm of my plots c2r1.south west,
    anchor=west,
] {(b)};
\node[below = 1.5cm of my plots c1r2.south west,
    anchor=west, xshift=4cm,
] {(c)};
\end{tikzpicture}
%
    \caption{POD basis training for Stokes flow:
    an example sample domain and solution for the flow-past-array problem (\ref{eq:stokes-array}),
    with parameter values of (a) $(g_1, g_2) = (0.5, 0.4)$ and (b) $(g_1, g_2) = (0.5, -0.4)$;
    and (c) the singular value spectra of the trained POD basis for the reference domains.
    The arrows indicate the flow velocity $\bu$.}
    \label{fig:stokes-sample}
\end{figure}
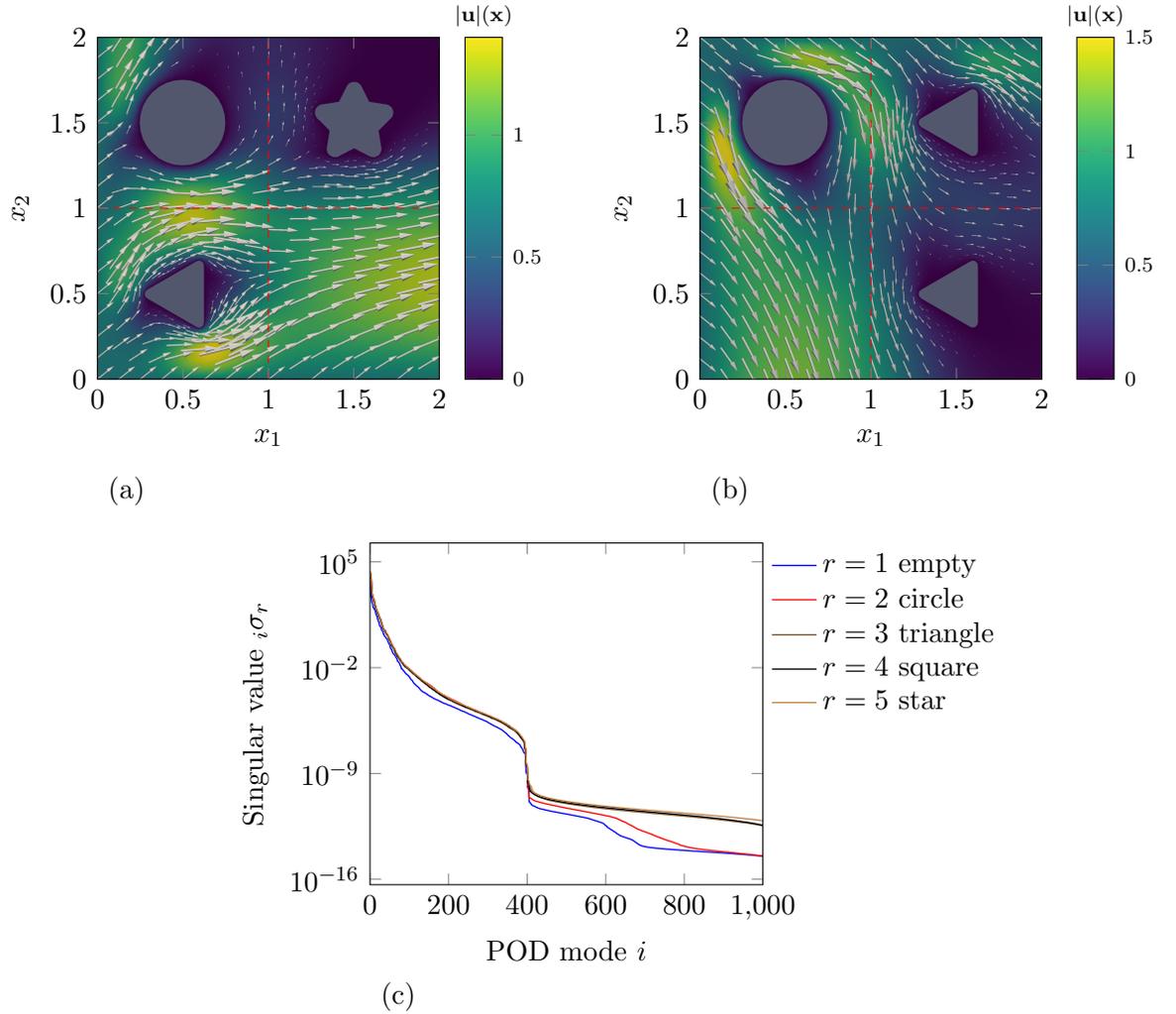

Sample snapshots for basis training are generated
on 2-by-2-component domains ${}_s\Omega$ for $s=1, 2, \ldots, 1400$.
For the $s$-th sample,
${}_s\Omega = \bigcup\limits_{m=1}^4 {}_s\Omega_m$ are constructed
with four randomly chosen subdomains ${}_s\Omega_m$ from $\rC$.
Figure~\ref{fig:stokes-sample}(a)~and~\ref{fig:stokes-sample}(b) show two example sample domains.
On each sample domain ${}_s\Omega$,
a random inflow velocity is set with a sinusoidal perturbation for the Dirichlet boundary condition,
\begin{subequations}\label{eq:stokes-array}
    \begin{equation}
        \bg_{di} = (g_1 + \Delta g_1\sin 2\pi(\mathbf{k}_1\cdot\bx + \theta_1), g_2 + \Delta g_2\sin 2\pi(\mathbf{k}_2\cdot\bx + \theta_2)),
    \end{equation}
    on the Dirichlet boundary ${}_s\partial\Omega_{di}$, which represents the upwind sides of ${}_s\Omega$.
    At the surface of the object inside the domain, a no-slip wall boundary is set,
    \begin{equation}
        \bg_{di} = \mathbf{0} \qquad \text{on } {}_s\partial\Omega_0 \equiv \left\{\bx\in {}_s\partial\Omega \;\big|\; x_1, x_2 \ne 0, 2 \right\}.
    \end{equation}
    The rest of the boundary is set as homogeneous Neumann boundary for ${}_s\partial\Omega_{ne} = {}_s\partial\Omega / ({}_s\partial\Omega_{di} \cup {}_s\partial\Omega_0)$,
    \begin{equation}
        \bg_{ne} = \mathbf{0} \qquad \text{on } {}_s\partial\Omega_{ne}.
    \end{equation}
\end{subequations}
Figure~\ref{fig:stokes-sample}~(a)~and~(b) show two example solutions from this flow-past-array problem (\ref{eq:stokes-array}).
The sample inflow velocity is chosen from a uniform distribution,
\begin{subequations}\label{eq:stokes-train-sample}
    \begin{equation}
        g_1, g_2 \in U[-1, 1]
    \end{equation}
    \begin{equation}
        \Delta g_1, \Delta g_2 \in U[-0.1, 0.1]
    \end{equation}
    \begin{equation}
        \mathbf{k}_1, \mathbf{k}_2 \in U[-0.5, 0.5]^2
    \end{equation}
    \begin{equation}
        \theta_1, \theta_2 \in U[0, 1].
    \end{equation}
\end{subequations}
Table~\ref{tab:stokes-comp} shows the summary of the number of sample snapshots collected on each reference domain.
Figure~\ref{fig:stokes-sample}(c) shows the singular value spectra of POD modes trained on the reference domains.
\par
In the trained POD modes, the velocity and pressure are coupled together.
In other words, the POD basis $\bPhi_r$ for the reference domain $r$ has the size of $(N_r, R_r)$ with $N_r = \dim(\bu_r) + \dim(p_r)$,
and its basis vectors represent the modes of $\bq_r = (\bu_r, p_r)$ as a whole.
The resulting ROM (\ref{eq:weak-gov-reduced}) is then composed of $M^2$ matrix blocks,
where its $(m, n)$ block matrix has the size of $(R_{r(m)}, R_{r(n)})$.

\subsection{Prediction performance}

\subsubsection{Global-level prediction}
\begin{figure}
    \begin{tikzpicture}[font=\small, spy using outlines={circle,black,magnification=6,size=1.5cm, connect spies}]
\pgfplotsset{set layers=standard}
    \begin{groupplot}[
        group style={
            group name = my plots,
            group size= 2 by 2,
            xlabels at =edge bottom,
            horizontal sep=3.5cm,
            vertical sep=2.5cm,
        },
        name=chung,
    ]    

        \nextgroupplot[
            height = 0.45\textwidth,
            width = 0.45\textwidth,
            ylabel={$x_2$},
            xlabel={$x_1$},
            tick scale binop ={\times},
            xmin = 0, xmax = 32,
            ymin = 0, ymax = 32,
            point meta min=0.0, point meta max=10.0,
            colorbar, colormap/viridis,
            colorbar style={
                font=\scriptsize,
                xticklabel pos=upper,
                scaled y ticks=false,
                /pgf/number format/precision=4,
                at={(rel axis cs: 1.01, 0.)}, anchor=south west,
                xlabel=$\vert \bu\vert(\bx)$,
            }
        ]
        
            \edef\imagepath{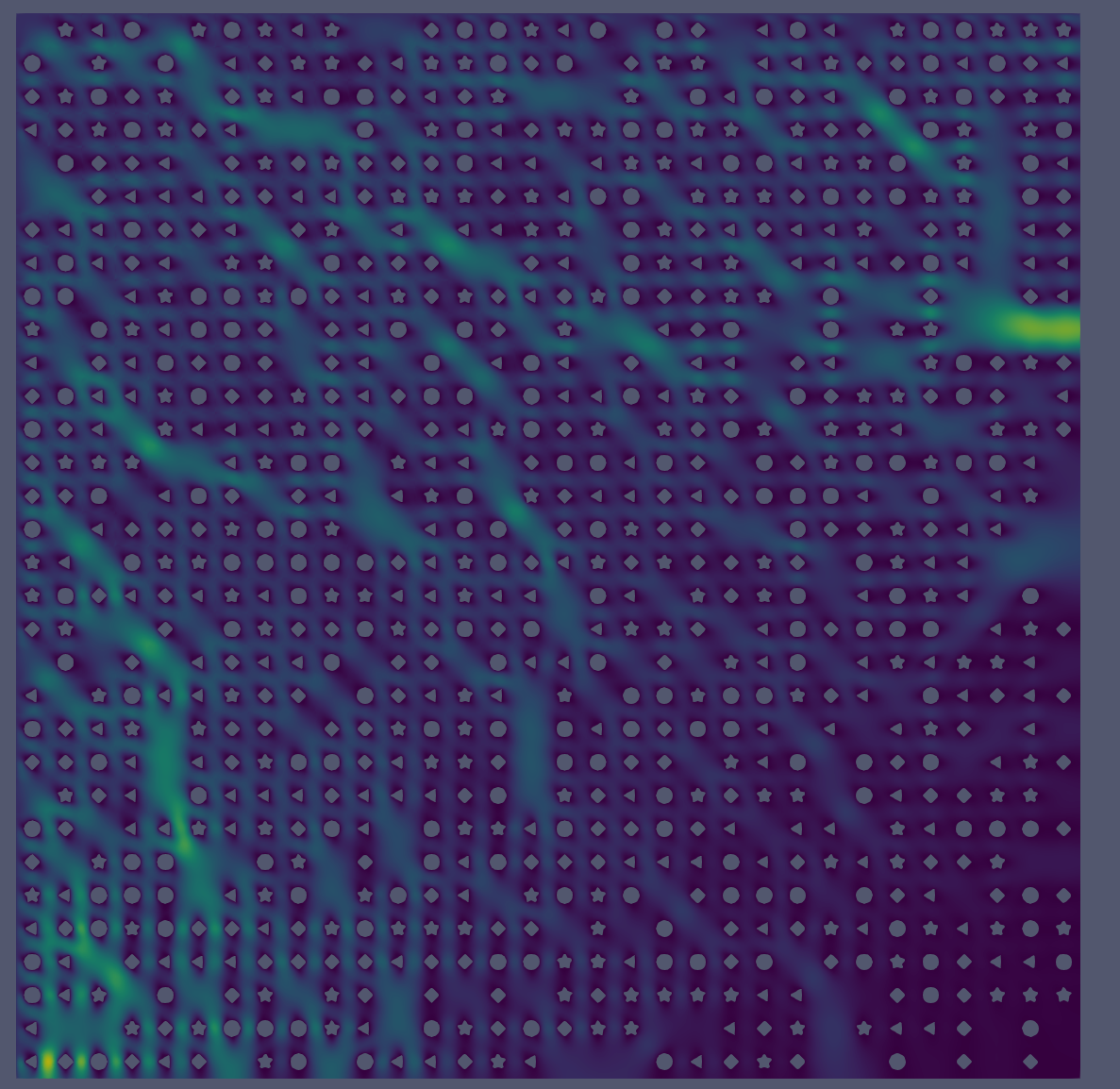}
            \addplot graphics[xmin=-0.5,xmax=34,ymin=-0.3,ymax=32.4]{\imagepath};

            \coordinate (pt) at (axis cs: 11.5, 23.);

        \nextgroupplot[
            height = 0.45\textwidth,
            width = 0.45\textwidth,
            ylabel={$x_2$},
            xlabel={$x_1$},
            tick scale binop ={\times},
            xmin = 0, xmax = 32,
            ymin = 0, ymax = 32,
            point meta min=0.0, point meta max=6.0,
            colorbar, colormap/viridis,
            colorbar style={
                font=\scriptsize,
                xticklabel pos=upper,
                scaled y ticks=false,
                /pgf/number format/precision=4,
                at={(rel axis cs: 2.8, 0.)}, anchor=south west,
                xlabel=$\vert \bu\vert(\bx)$,
            }
        ]
        
            \edef\imagepath{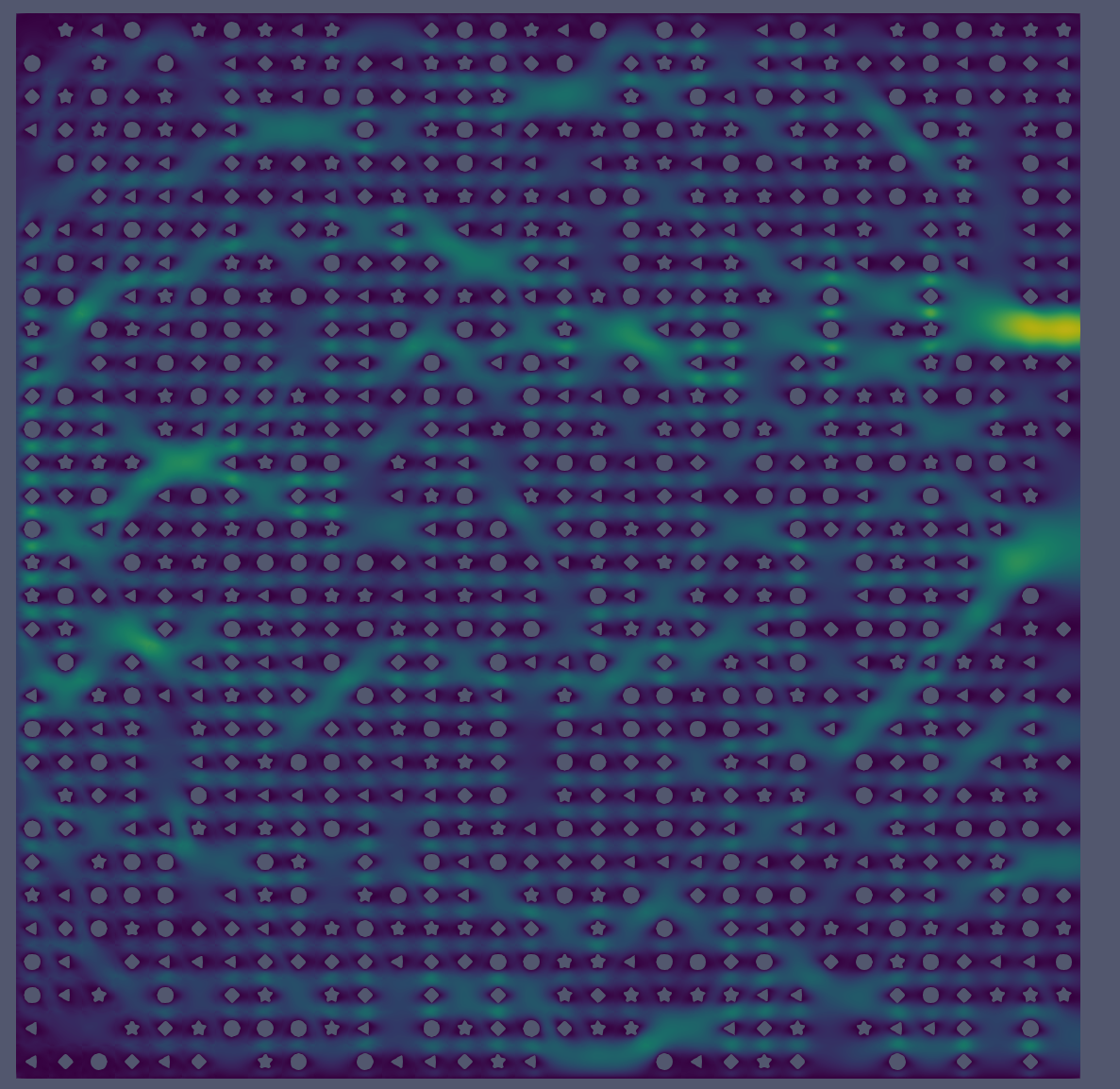}
            \addplot graphics[xmin=-0.5,xmax=34,ymin=-0.3,ymax=32.4]{\imagepath};

\end{groupplot}
\node[pin={[pin edge={red,thick},pin distance=1.2cm]90:{%
    \begin{tikzpicture}[baseline,trim axis left,trim axis right]
        \begin{axis}[
            height=3cm,
            width=3cm,
            xtick = \empty,
            ytick = \empty,
            xmin = 10.2, xmax=12.2,
            ymin = 23., ymax=25.,
            axis line style={black,
                                    line width=1.5},
        ]
            \edef\imagepath{./array_vel_mag.png}
            \addplot graphics[xmin=-0.5,xmax=34,ymin=-0.3,ymax=32.4]{\imagepath};
            
        \end{axis}
    \end{tikzpicture}%
}}] at (pt) {};

\node[below = 1.5cm of my plots c1r1.south west,
    anchor=west,
] {(a)};
\node[below = 1.5cm of my plots c2r1.south west,
    anchor=west,
] {(b)};
\end{tikzpicture}
%
    \caption{Test ROM predictions on $\Omega(32)$:
    (a) flow past array (\ref{eq:stokes-array});
    and (b) channel flow (\ref{eq:stokes-channel}).}
    \label{fig:stokes-global}
\end{figure}
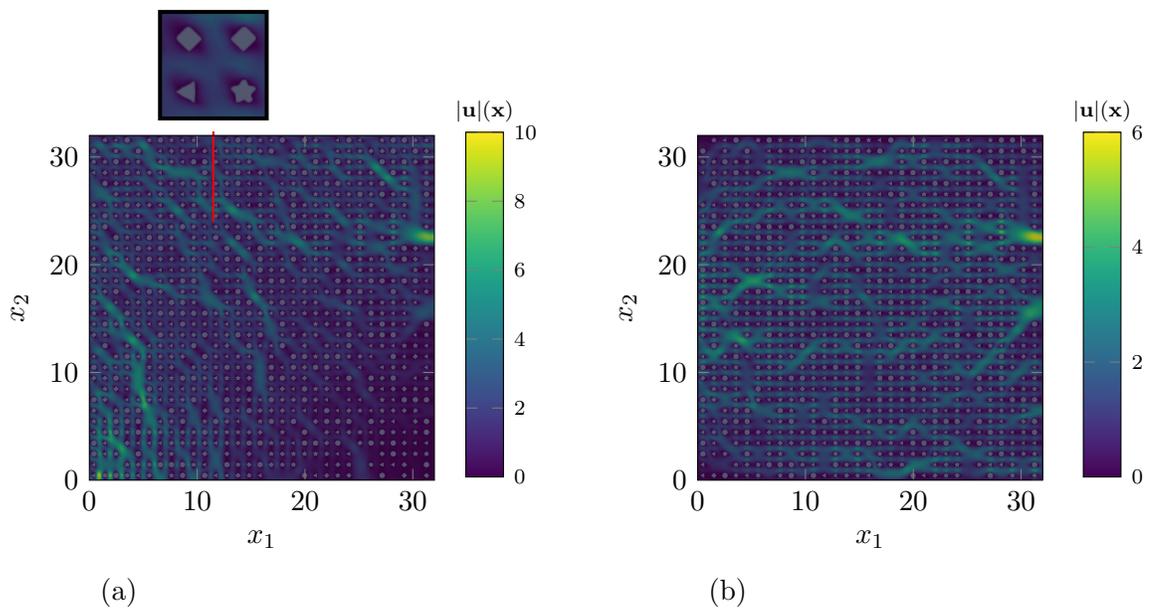
We demonstrate the performance of the ROM with the same sample boundary condition (\ref{eq:stokes-array}),
but on global domain $\Omega(N_c) \subset [0, N_c]^2$ filled with $N_c^2$ random components.
The inflow velocity $\bg_{di}$ for (\ref{eq:stokes-array}) is chosen from an extended uniform distribution for 100 test cases at each size of the domain,
\begin{equation}
    (g_1, g_2) \in U[-1.5, 1.5]^2.
\end{equation}
Figure~\ref{fig:stokes-global}(a) shows the prediction made by ROM on a $\Omega(32)$.
\par
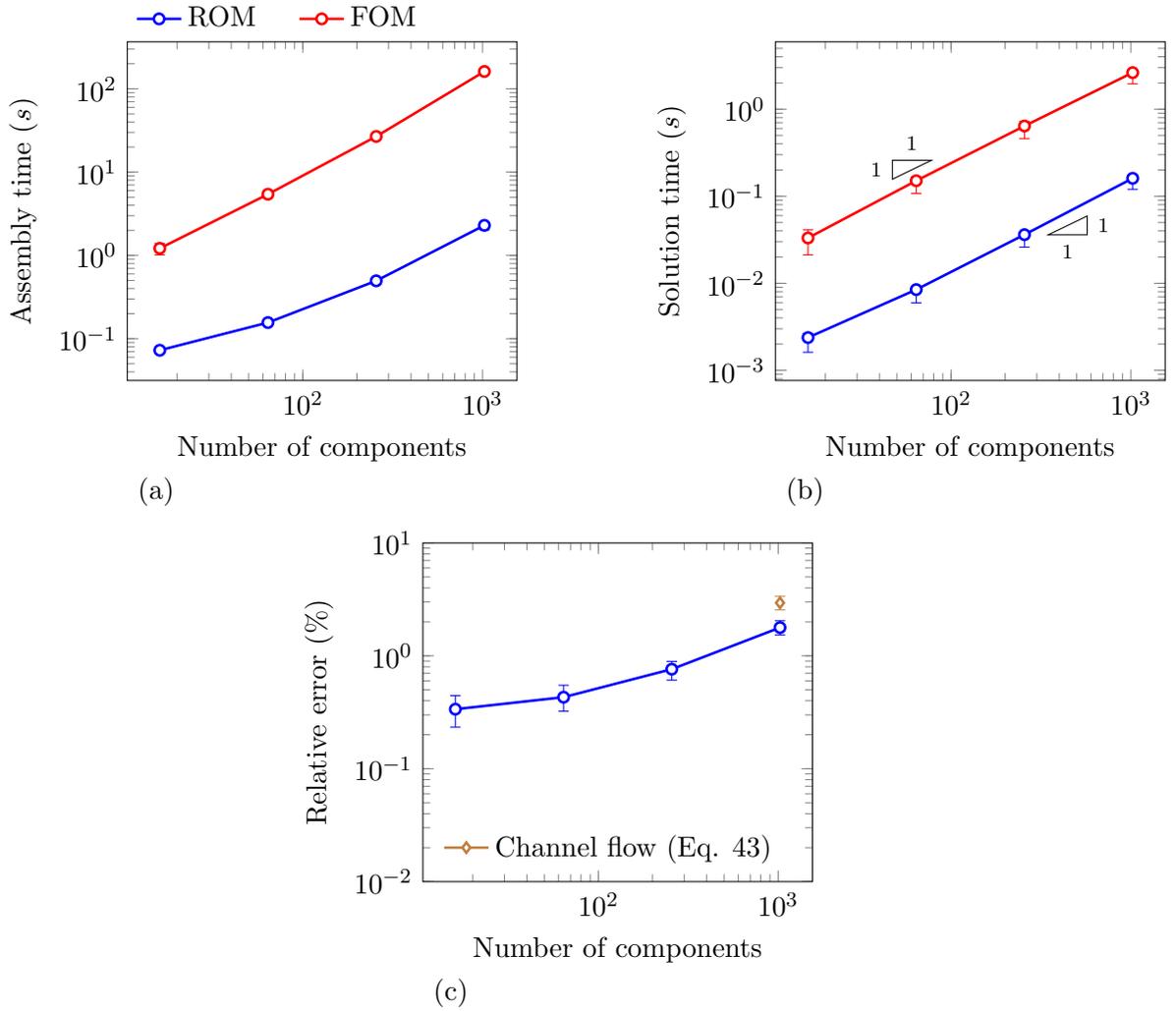
\begin{figure}
    \begin{tikzpicture}[font=\small,]
    \begin{groupplot}[
        group style={
            group name = my plots,
            group size= 2 by 2,
            xlabels at =edge bottom,
            horizontal sep=3.5cm,
            vertical sep=2.2cm,
        },
        name=chung,
    ]    
\pgfplotsset{set layers=standard}%

        \nextgroupplot[
            height = 0.45\textwidth,
            width = 0.5\textwidth,
            xlabel={Number of components},
            ylabel={Assembly time ($s$)},
            tick scale binop ={\times},
            xmode=log, ymode=log,
            legend style={
                font=\small,
                draw=none, fill=none,
                at={(rel axis cs: 0., 1.0)},
                anchor=south west,
                nodes={scale=1.0},
                legend cell align={left},
                legend columns=4,
                /tikz/every even column/.append style={column sep=0.5cm},
            },
            legend image post style={mark options={scale=1.0, fill=white, line width=1.0}},
        ]
        




            \addplot+ [
                line width=1.0,
                solid,
                mark=*,
                mark options={fill=white,},
                blue,
                error bars/.cd, y dir=both, y explicit,
            ]
            table [
                x expr=\thisrowno{0}^2, y index=1,
                y error minus expr=\thisrowno{1} - \thisrowno{4},
                y error plus expr=\thisrowno{5} - \thisrowno{1},
            ]{./stokes_direct_rom_assemble.txt};

            \addplot+ [
                line width=1.0,
                solid,
                mark=*,
                mark options={fill=white,},
                red,
                error bars/.cd, y dir=both, y explicit,
            ]
            table [
                x expr=\thisrowno{0}^2, y index=1,
                y error minus expr=\thisrowno{1} - \thisrowno{4},
                y error plus expr=\thisrowno{5} - \thisrowno{1},
            ]{./stokes_direct_fom_assemble.txt};


            \legend{ROM, FOM}

        \nextgroupplot[
            height = 0.45\textwidth,
            width = 0.5\textwidth,
            xlabel={Number of components},
            ylabel={Solution time ($s$)},
            tick scale binop ={\times},
            xmode=log, ymode=log,
        ]
        




            \addplot+ [
                line width=1.0,
                solid,
                mark=*,
                mark options={fill=white, solid,},
                blue,
                error bars/.cd, y dir=both, y explicit,
            ]
            table [
                x expr=\thisrowno{0}^2, y index=1,
                y error minus expr=\thisrowno{1} - \thisrowno{4},
                y error plus expr=\thisrowno{5} - \thisrowno{1},
            ]{./stokes_direct_rom_solve.txt};

            \addplot+ [
                line width=1.0,
                solid,
                mark=*,
                mark options={fill=white, solid,},
                red,
                error bars/.cd, y dir=both, y explicit,
            ]
            table [
                x expr=\thisrowno{0}^2, y index=1,
                y error minus expr=\thisrowno{1} - \thisrowno{4},
                y error plus expr=\thisrowno{5} - \thisrowno{1},    
            ]{./stokes_direct_fom_solve.txt};

            \logLogReverseSlopeTriangle{0.3}{0.1}{0.65}{1}{font=\scriptsize}{south}
            \logLogSlopeTriangle{0.8}{0.1}{0.43}{1}{font=\scriptsize}{north}
 
        \nextgroupplot[
            height = 0.45\textwidth,
            width = 0.5\textwidth,
            xlabel={Number of components},
            ylabel={Relative error ($\%$)},
            tick scale binop ={\times},
            xmode=log, ymode =log,
            ymin=1e-2, ymax=1e1,
            xshift=4cm,
            legend pos=south west,
            legend style={
                font=\small,
                draw=none, fill=none,
                anchor=south west,
                nodes={scale=1.0},
                legend cell align={left},
            },
        ]

            \addlegendimage{
                line width=1.0,
                draw=none,
                mark=diamond*,
                mark options={fill=white,},
                brown,
            }
            \addlegendentry{Channel flow (Eq. \ref{eq:stokes-channel})}
        

            \addplot+ [
                line width=1.0,
                solid,
                mark=*,
                mark options={fill=white,},
                blue,
                error bars/.cd, y dir=both, y explicit,
            ]
            table [
                x expr=\thisrowno{0}^2, y expr=\thisrowno{1} * 1e2,
                y error minus expr=(\thisrowno{1} - \thisrowno{4}) * 1e2,
                y error plus expr=(\thisrowno{5} - \thisrowno{1}) * 1e2,
            ]{./stokes_direct_rel_error.txt};

            \addplot+ [
                line width=1.0,
                solid,
                mark=diamond*,
                mark options={fill=white,},
                brown,
                error bars/.cd, y dir=both, y explicit,
            ]
            table [
                x expr=\thisrowno{0}^2, y expr=\thisrowno{1} * 1e2,
                y error minus expr=(\thisrowno{1} - \thisrowno{4}) * 1e2,
                y error plus expr=(\thisrowno{5} - \thisrowno{1}) * 1e2,
            ]{./stokes_channel_rel_error.txt};

  \end{groupplot}
\node[below = 1.5cm of my plots c1r1.south west,
    anchor=west,
] {(a)};
\node[below = 1.5cm of my plots c2r1.south west,
    anchor=west,
] {(b)};
\node[below = 1.5cm of my plots c1r2.south west,
    anchor=west, xshift=4cm,
] {(c)};
\end{tikzpicture}
%
    \caption{Performance of the ROM compared to the FOM for the Stokes flow equation, depending on the number of components $M$:
    (a) assembly time of the system;
    (b) computation time of the system; and
    (c) relative error of the ROM compared to the FOM solution.
    The marker denotes the median value of 100 test cases,
    and the error bar denotes $95\%$-confidence interval of the test cases.
    }
    \label{stokes-scaleup}
\end{figure}
The ROM constructed for the Stokes flow was capable of robust \PR{extrapolation} in all scales,
similar to the Poisson equation shown in Section~\ref{sec:poisson}.
\todo{KC: should switch the order of subfigures in Figure~\ref{stokes-scaleup}?}
Figure~\ref{stokes-scaleup}~(c) shows
the relative error of the ROM over 100 test cases at each size of the domain,
which remains \PR{between} $0.3\%$ to $2\%$.
Unlike for the Poisson equation, the variance between test cases is not large,
slowly increasing with the size of the domain.
This increasing error seems related to the extrapolative behavior emerging on a larger scale of domain.
In example solutions in Figure~\ref{fig:stokes-global},
\TYL{we observe that the flow tends to accumulate in the `empty' components scattered in the domain,
corresponding to the regions of least resistance,}
leading to \PR{absolute} velocity \PR{much higher than} the training data.
The maximum velocity magnitude reaches up to $\max(\vert\bu\vert)\sim 10$,
which is an order of magnitude larger than the training range (\ref{eq:stokes-train-sample}).
Even for such a strongly extrapolative case,
the ROM achieved a robust prediction with \PR{only} a relative error of $\sim 2\%$.
\par
Such robust ROM prediction is also made much faster compared to FOM.
As shown in Figure~\ref{stokes-scaleup}~(a),
the ROM accelerates the assembly time for $M=16$ about 10 times faster,
and achieves a better scaling with $M$.
Also, in Figure~\ref{stokes-scaleup}~(b),
the computation time is reduced by a factor of 15 over all sizes of the domain.
\par
In order to demonstrate robust prediction of ROM outside the training range,
we also made ROM predictions on a channel flow boundary condition, \todo[inline]{PR: Should we mention the Reynolds number for channel flow?\\ KC: at this point, I think mentioning Reynolds number is somewhat irrelevant.}
\begin{subequations}\label{eq:stokes-channel}
    \begin{equation}
        \bg_{di} =
        \begin{cases}
            \left( U_{in}\left[1 - 4\left(\frac{x_2}{N_c} - \frac{1}{2}\right)^2\right], 0 \right) & \text{on } x_1 = 0, \\
            \mathbf{0} & \text{on } x_2 = 0, N_c. \\
        \end{cases}
    \end{equation}
    Note that the samples generated with the flow-past-array problem (\ref{eq:stokes-array})
    never have the no-slip wall condition on their outside boundaries (${}_s\partial\Omega_{di}$).
    This channel flow condition therefore poses a problem that is qualitatively different from the training data.
    \TYL{Additionally, note that physically, the channel flow inlet boundary condition should occur at a slightly
        greater distance upstream, but this does not affect the present discussion.}
    On the outflow direction boundary, the homogeneous Neumann boundary condition is applied,
    \begin{equation}
        \bg_{ne} = \mathbf{0} \qquad\text{on } x_1 = N_c.
    \end{equation}
    On the object surfaces inside the domain, no-slip wall boundary is set,
    \begin{equation}
        \bg_{di} = \mathbf{0} \qquad\text{on } \left\{ \bx\in\partial\Omega(N_c) \;\big|\; x_1, x_2 \ne 0, N_c \right\}.
    \end{equation}
\end{subequations}
A test predictive ROM solution is shown in Figure~\ref{fig:stokes-global}~(b).
In Figure~\ref{stokes-scaleup}~(c),
the global ROM constructed from the flow-past-array problem (\ref{eq:stokes-array})
made similarly robust predictions with relative error of $\sim3\%$,
only slightly larger than the case of (\ref{eq:stokes-array}).
The assembly and solving time did not show a difference in this case.
This demonstrates the robustness of the component ROM against general problems under the same physics.

\subsubsection{Performance scaling}
As for the Poisson equation in Section~\ref{subsubsec:poisson-scaling},
the performance scaling with the number of basis is investigated for the Stokes flow equation.
The solution time and the relative error of the ROM are measured
for the flow past array problem (\ref{eq:stokes-array}) on a random configuration of size $\bOmega(4)$.
\begin{figure}[tbp]
    \begin{tikzpicture}[font=\small,]
    \begin{groupplot}[
        group style={
            group name = my plots,
            group size= 2 by 1,
            xlabels at =edge bottom,
            horizontal sep=3.5cm,
            vertical sep=2.5cm,
        },
        name=chung,
    ]    
\pgfplotsset{set layers=standard}%

        \nextgroupplot[
            height = 0.4\textwidth,
            width = 0.5\textwidth,
            xlabel={Number of basis},
            ylabel={Solution time ($s$)},
            tick scale binop ={\times},
            xmode=log, ymode=log,
            xmax=1e3,
            legend style={
                font=\small,
                draw=none, fill=none,
                at={(rel axis cs: 0., 1.0)},
                anchor=south west,
                nodes={scale=1.0},
                legend cell align={left},
                legend columns=4,
                /tikz/every even column/.append style={column sep=0.5cm},
            },
            legend image post style={mark options={scale=1.0, fill=white, line width=1.0}},
        ]
        

            \addplot+ [
                line width=1.0,
                solid,
                mark=*,
                mark options={fill=white,},
                blue,
            ]
            table [x index=0, y index=2]{./stokes_array_4x4.direct.txt};


            \draw[red, solid,] (axis cs: 1e0, 3.43427E-02) -- (axis cs: 1e4, 3.43427E-02);
            \node[anchor=north west, font=\scriptsize, xshift=10pt, red] at (axis cs: 1e1, 3.43427E-02) {FOM};

            \logLogSlopeTriangle{0.4}{0.1}{0.27}{1}{}{north}

        \nextgroupplot[
            height = 0.4\textwidth,
            width = 0.5\textwidth,
            xlabel={Number of basis},
            ylabel={Relative error ($\%$)},
            tick scale binop ={\times},
            xmode=log, ymode=log,
            xmin=1e1, ymax=2e2,
            legend style={
                font=\small,
                draw=none, fill=none,
                at={(rel axis cs: 0., 1.0)},
                anchor=south west,
                nodes={scale=1.0},
                legend cell align={left},
                legend columns=4,
                /tikz/every even column/.append style={column sep=0.5cm},
            },
            legend image post style={mark options={scale=1.0, fill=white, line width=1.0}},
        ]
        

            \addplot+ [
                line width=1.0,
                solid,
                mark=*,
                mark options={fill=white,},
                blue,
                select coords between index={0}{4},
            ]
            table [x index=0, y expr=\thisrowno{5} * 1e2]{./stokes_array_4x4.direct.txt};

    
            \logLogSlopeTriangle{0.37}{0.1}{0.82}{-8}{}{south}
 
\end{groupplot}
\node[below = 1.5cm of my plots c1r1.south west,
    anchor=west,
] {(a)};
\node[below = 1.5cm of my plots c2r1.south west,
    anchor=west,
] {(b)};
\end{tikzpicture}
%
    \caption{Performance of the component ROM for the Stokes flow equation:
    (a) solution time;
    and (b) relative error of the ROM solution.
    The solution time of the FOM is plotted (red) for reference.}
    \label{fig:stokes-scaling}
\end{figure}
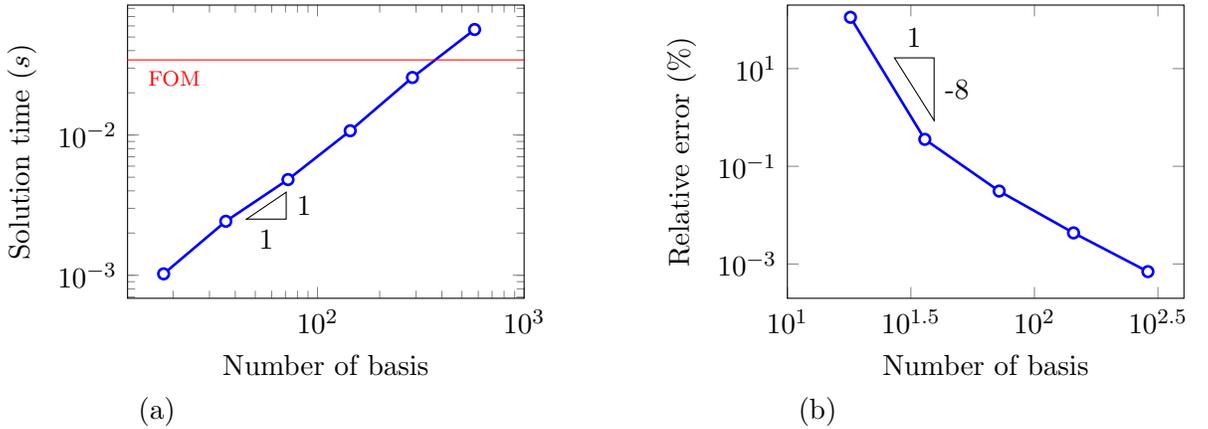
In Figure~\ref{fig:stokes-scaling},
similar to the Poisson equation,
the accuracy of the ROM scales rapidly with the number of basis.
With only $R_r=36$ basis, the ROM already achieves $\sim0.36\%$ relative error,
while accelerating the solution time by a factor of $15$.
This again demonstrates the effectiveness of the ROM with the low-dimensional linear subspace.
As for the Poisson equation,
the scaling slows down as we increase $R_r$,
following the singular value spectra in Figure~\ref{fig:stokes-sample}(c).

\section{Conclusion}
In this work we address a common practical challenge in scaling up,
where
the data for the model reduction is provided only at small component-level scales,
and accurate prediction is required at a much larger scale.
We developed a component ROM where the ROM is combined with DG-DD.
The proposed method provides building blocks for robust and efficient prediction at an extrapolated scale.
\par
The proposed method is demonstrated on two linear PDEs: the Poisson equation and the Stokes flow equation.
For both cases, component ROM accelerates the solving time by a factor of $15\sim40$,
for global domains up to $1000$ \PR{times} larger than the component domains.
The resulting ROM was capable of robust prediction,
with relative error of $\sim3\%$ over all scales,
even for the problems that are qualitatively different from the samples used for the linear subspace training.
The accuracy of the ROM scales with the linear subspace dimension $R$
as $\cO(R^{-5})$ for Poisson and as $\cO(R^{-8})$ for Stokes.
This is much faster than the $\cO(R)$ scaling of the corresponding solution time,
enabling the computation to be both efficient and accurate with a low dimensional linear subspace.
\par
Even for the ROM, of course, the computational cost scales with the size of the problem.
Parallel computation of the ROM will be necessary for larger scale problems, particularly for three-dimensional simulations.
Though not demonstrated here, our approach is not limited to the spatial dimension or the serial computing platform.
Preconditioners for the component ROM, once developed, can achieve further acceleration at the larger scale problem.
For positive-definite systems like he Poisson equation,
two-level Schwarz preconditioners can be promising~\cite{Antonietti2016}.
For indefinite systems like the Stokes flow equation,
block preconditioners might need further factorization to maintain positive-definiteness.
\par
It is well known that the linear subspace ROM cannot be naively applied to nonlinear problems.
For polynomially nonlinear PDEs, such as the incompressible Navier--Stokes equation,
the tensorial approach can be applied~\cite{Lassila2014}.
In other cases, hyper-reduction approaches are also available~\cite{Willcox2006,Chaturantabut2010,Yano2019}.
In all cases, similar component ROM variants can be formulated with the DG-DD,
as all the model reduction techniques similarly exploit the physics equation.

\section*{Acknowledgement}

This work was performed under the auspices of the U.S. Department of Energy
by Lawrence Livermore National Laboratory under contract DE-AC52-07NA27344
and was supported by Laboratory Directed Research and Development funding under project 22-SI-006.
LLNL-JRNL-857774.




\bibliographystyle{elsarticle-num} 
\bibliography{references}

%
%
%
%
\end{document}